\documentclass[onecolumn]{IEEEtran}
\usepackage{times}

\usepackage{amsmath}  
\usepackage{amssymb}  
\usepackage{mathrsfs} 
\usepackage{algorithm,algorithmic}

\usepackage{theorem}  
\usepackage{cite}     
\usepackage{comment}  

\usepackage{upref}
\usepackage{amsfonts}

\usepackage{verbatim}

\usepackage[dvipsnames,usenames]{color}
\usepackage{enumerate}

\usepackage{graphicx}
\usepackage{subfigure}

\usepackage{latexsym}
\usepackage[normalem]{ulem}

\newcommand{\be}[1]{\begin{equation}\label{#1}}
\newcommand{\ee}{\end{equation}}

\newcommand{\bc}{\begin{center}}
\newcommand{\ec}{\end{center}}

\newcommand{\cC}{{\cal C}}

\newcommand{\cE}{{\cal E}}

\newcommand{\cR}{{\cal R}}

\newcommand{\cU}{{\cal U}}

\newcommand{\cX}{{\cal X}}

\newcommand{\bfv}{{\boldsymbol v}}

\newcommand{\bfx}{{\boldsymbol x}}
\newcommand{\bfy}{{\boldsymbol y}}
\newcommand{\bfz}{{\boldsymbol z}}

\renewcommand{\leq}{\leqslant}

\renewcommand{\geq}{\geqslant}

\newcommand{\Cref}[1]{Co\-rol\-la\-ry\,\ref{#1}}

\newcommand{\nchoosek}[2]{\left(\begin{array}{c}#1\\#2\end{array}\right)}

\theoremstyle{plain} \theorembodyfont{\normalfont\slshape}

\newtheorem{thm}{Theorem$\!$}
\newenvironment{theorem}{\begin{thm}\hspace*{-1ex}{\bf.}}{\end{thm}}

\newtheorem{prop}[thm]{Proposition$\!$}
\newenvironment{proposition}{\begin{prop}\hspace*{-1ex}{\bf.}}{\end{prop}}

\newtheorem{lem}[thm]{Lemma$\!$}
\newenvironment{lemma}{\begin{lem}\hspace*{-1ex}{\bf.}}{\end{lem}}

\newtheorem{cor}[thm]{Corollary$\!$}
\newenvironment{corollary}{\begin{cor}\hspace*{-1ex}{\bf.}}{\end{cor}}

\newtheorem{prob}[thm]{Problem$\!$}

\newtheorem{defi}[thm]{Definition$\!$}

\theorembodyfont{\normalfont}

\newtheorem{exam}{Example$\!$}
\newenvironment{example}{\begin{exam}\hspace*{-1ex}{\bf .}}{\end{exam}}

\newtheorem{remrk}{Remark$\!$}

\definecolor{Codecolor}{named}{White}  

\newcommand{\Copen}{\mbox{\{\kern-5.50pt\{}}
\newcommand{\Cclose}{\mbox{\}\kern-5.50pt\}}}
\newcommand{\Cslash}{\mbox{$\backslash\kern-6.02pt\backslash$}}

\begin{document}
\title{Unique Reconstruction of Coded Strings from Multiset Substring Spectra}
\author{
  \IEEEauthorblockN{
    Ryan~Gabrys}~and
    Olgica~Milenkovic\\
  {\normalsize
    \begin{tabular}{ccc}
     ECE Department, University of Illinois, Urbana-Champaign \\
    \end{tabular}}\vspace{-3ex}
   \vspace{-0.05in}
    }
\maketitle
\begin{abstract} The problem of reconstructing strings from their substring spectra has a long history and in its most simple incarnation asks for determining under which conditions the spectrum uniquely determines the string. 
We study the problem of \emph{coded string reconstruction} from multiset substring spectra, where the strings are restricted to lie in some codebook. In particular, we consider binary codebooks that allow for unique string reconstruction and propose a new method, termed \emph{repeat replacement}, to create the codebook. Our contributions include algorithmic solutions for repeat replacement and constructive redundancy bounds for the underlying coding schemes. We also consider extensions of the problem to noisy settings in which substrings are compromised by burst and random errors. The study is motivated by applications in DNA-based data storage systems that use high throughput readout sequencers.
\end{abstract}

\vspace{-0.2in}
\section{Introduction}

String reconstruction refers to the problem of recovering a string based on some information about {its substrings (i.e, strings composed of consecutive elements of the string) or subsequences (i.e, strings composed of possibly non-consecutive elements of the string).} Most often, the information is in the form of the sequence spectrum~\cite{MarSki95}, comprising all distinct substrings of the string; the multiset spectrum~\cite{kiah2016}, comprising the multiset of the substrings of the string; the $k$-deck~\cite{Scott,MMSSS91}, comprising all subsequences of the string of length $k$; sequence traces~\cite{DS02,batu2004}, corresponding to randomly selected subsequences of the string; or multiset compositions~\cite{acharya2015,acharya2017}, providing information about the composition of substrings only. The central problem in string reconstruction is to identify which strings may be uniquely reconstructed given a certain substring and subsequence information. In all of the above described scenarios, no restrictions are imposed on the properties of the strings one seeks to reconstruct.

In contrast, the problem of coded string reconstruction asks for reconstructing strings that satisfy some predefined constraints. The first instance of a coded sequence reconstruction problem was studied by Levenshtein~\cite{levenshtein}, who posed the sequence reconstruction problem for strings drawn from an error-correcting codebook. Recently, a new form of coded reconstruction was introduced in~\cite{kiah2016,CCEK17,GM17}, with the goal of performing string encodings that enable unique reconstruction based on substring multisets. The problem of interest is to identify efficient coding schemes that convert arbitrary input strings into strings that may be uniquely reconstructed given some predetermined substring and/or subsequence information. It is for this setting that we propose a new coding method termed \emph{repeat replacement}. Repeat replacement may be viewed as a form of string compression in which one removes all repeated substrings of prescribed length and replaces them with pointers to their original locations in a way that allows for recovering the original information. In many cases, the proposed techniques require careful selection of the order of repeat removals and involve a special encoding process for the repeats. Other replacement techniques were investigated in~\cite{WiImm10,LY17}, with the goal of imposing runlength or balancing constraints on a string. 
In these scenarios, removing offending substrings does not cause the introduction of other offending substrings, which makes the underlying problem solution simpler than repeat replacement. Another important observation is that classical compression algorithms such as Lempel-Ziv (LZ)-type methods~\cite{lempelziv} cannot be used instead of repeat replacement: It is straightforward to see that there is no guarantee that single-pass LZ encoding removes all repeats and does not introduce new $1$s in the compressed string~\cite{szpankowski}. 

{The problem of coded reconstruction analyzed in this work is motivated by applications in DNA-based data storage, where the strings to be sequenced are user-defined and synthetically generated, and hence allowed to have arbitrary content~\cite{goldman,grass}. The first DNA-based storage system implementation used 
suffix-prefix overlapping DNA-blocks, termed oligos, to store information and ensure four-fold coverage of each information symbol~\cite{goldman}. Such a representation does not allow for random access, extensive error-correction or rewriting. To address these issues, the authors of~\cite{yazdi2015,yazdi2016,yazdiWMU} proposed using long DNA blocks (gBlocks), and showed how these blocks can be equipped with addresses that enable random access and sequenced with nanopores. For long block lengths, no overlap between the blocks is required, and reconstruction of the original message is accomplishes by individually decoding of all blocks. A related line of coding problems was reported in~\cite{LWZY18,Sima}. 
It is in the context of the original system architecture~\cite{goldman} that the issue of substrings repeats is of importance. It is known that certain repeated substrings lead to ambiguities in string reconstruction and cause assembly errors~\cite{ukkonen1992} independently on the sequence coverage, whenever the observed substring length is below some critical threshold. Hence, to enable unique reconstruction and reduce the critical threshold length, one has to design codebooks of strings that do not contain undesirable repeats. The main finding of this work is that  \emph{even two bits of coding redundancy} allow the critical length to be reduced from $O(n)$ to $O(\log\,n)$, where $n$ denotes the length of the string to be stored. The redundancy is added in the process of repeat replacement, a procedure that requires $O(n^2 \log n)$ operations. The process of repeat replacement is generalized for the case when not all substrings of the string are observed (i.e., for the case of \emph{coverage errors}, which may be random or bursty), provided that the gap between observed reads is bounded by some constant independent {of} $n$. In addition, parallel findings are presented for additive substring errors accompanied by coverage errors.} 

{We also observe that the reconstruction problem simplifies as the alphabet size increases. In particular, a simple mapping akin to the one reported in~\cite{yazdiWMU} may be used to design reconstruction codes over non-binary alphabets using reconstruction codes for binary alphabets. Consequently, we focus our attention on the problem of designing reconstruction codes for binary alphabets only.}

This paper is organized as follows. Section~\ref{sec:notation} introduces the relevant notation and provides a rigorous problem statement. The section also discusses prior results in the area. Section~\ref{sec:g2logn} describes the codebook design process for substrings of length $> 2 \log n$ observed in a noiseless manner. Section~\ref{sec:l2logn} focuses on the case of string reconstruction with substring lengths within the interval $(\log n, 2 \log n)$. Section~\ref{sec:errors} studies the coded string recovery problem for noisy substring spectra. 

\section{Notation and Preliminaries} \label{sec:notation}

We consider the problem of reconstructing  a string $\bfx \in \{0,1\}^n$ from the multiset of its substrings of length $L$, where $L$ is allowed to scale with the string length $n$. Adopting a similar terminology as the one used in~\cite{MarSki95}, we let $\bfx_{i,L}$ denote the $i$-th substring in $\bfx$ of length $L$, i.e., the substring of $\bfx$ starting at position $i$. Furthermore, for $\bfx \in \{0,1\}^n$, we say that $\bfx_{1,L}$ is a prefix of $\bfx$ of length $L$ and that $\bfx_{n-L+1,L}$ is a suffix of $\bfx$ of length $L$. If $L\neq n$, then $\bfx_{n-L+1,L}$ is called a proper suffix of $\bfx$. For example, if $\bfx = (0,1,1,0)$, then $\bfx_{3,2} = (1,0)$. We refer to the multiset of all substrings of length $L$ of $\bfx$ as the $L$-multispectrum {(or substring profile)} of $\bfx$, and denote the it by $M_L(\bfx)$. 
Clearly, $M_L(\bfx)=\{{\bfx_{1,L},\bfx_{2,L},\ldots,\bfx_{n-L+1,L}\}}$. 

The period of a string $\bfx \in \{0,1\}^n$ is the smallest integer $p$ such that for all $i \in [n-p]$, $x_i = x_{i+p}$. For example, the string $\bfx = (0,1,1,0,1,1)$ of length $n=6$ has period $p=3$; we use $p(\bfx)$ to denote the period of $\bfx$ and refer to a string $\bfx$ with $p(\bfx)=n$ as aperiodic. The following example illustrates the previously introduced concepts.

\begin{example} Let $\bfx = (0,0,1,0,0)$. For $L=2$, $M_2(\bfx) = \Big \{ (0,0), (0,1), (1,0), (0,0) \Big \}$. The string $\bfx$ has period $3$.

Next, consider two strings $\bfx =(0,1,1,0,1)$ and $\bfy = (1,1,0,1,1)$. Clearly,
$$ M_3(\bfx) = \Big \{ (0,1,1), (1,1,0), (1,0,1) \Big \} = M_3(\bfy), $$
so that $\bfx$ and $\bfy$ cannot be reconstructed uniquely from their $3$-multispectra. However, for $L=4$,
$$ M_4(\bfx) = \Big \{ (0,1,1,0), (1,1,0,1) \Big \}, $$
and
$$ M_4(\bfy) = \Big \{ (1,1,0,1), (1,0,1,1) \Big \}, $$
and since $M_4(\bfx) \neq M_4(\bfy)$, $\bfx$ and $\bfy$ can be distinguished based on their $4$-multispectra.
\end{example}


In what follows, we will be concerned with constructing codebooks $\cX$ of strings such that all strings $\bfx \in \cX$ can be uniquely determined by their $L$-multispectra $M_L(\bfx)$. Such codebooks are referred to as \textit{$L$-reconstruction codes}. We define the rate of a codebook $\cX \subseteq \{0,1\}^n$ as $\frac{ \log |\cX|}{n}$, where the $\log$ is base two, and the codebook redundancy equals $n- \log |\cX|$. The maximum rate of any $L$-reconstruction code of length $n$ is denoted by $R(n,L)$. The goal of this work is to design $L$-reconstruction codes with smallest redundancy, and consequently, largest rate. Another requirement is that one should be able to perform message encoding and decoding of such codes in a straightforward and computationally efficient manner. 

{Following up on the work in~\cite{kiah2016}, the authors of~\cite{CCEK17} showed that for $L = \frac{n}{2} + 1$, there exists an $L$-reconstruction code with one bit of redundancy, endowed with a simple encoding scheme\footnote{Here, and elsewhere in the paper, we tacitly assume that the values of $L$ and $\log\,n$ are integers. In general, one needs to use floor functions for each fractional value. To avoid notational clutter, we dispose of the floor function and only write the corresponding function of $n$.}. In addition, the same paper established that for $L \geq 2 \log n + 2,$ it is possible to design a codebook with $\frac{2^{n-1}}{n}$ uniquely recoverable strings. As a consequence, for this parameter regime, one has
$$ \lim_{n \to \infty} R(n,L) \to 1,$$
whenever $L \geq 2 \log n + 2$. No explicit encoding schemes are known for codes that achieve this rate.} 

{As noted in \cite{CCEK17}, $| \{ M : \exists \, \bfx \in \{0,1\}^n, M_L(\bfx) = M  \}|$ is at most equal to the number of $2^L$-compositions of $n-L+1$, so that
\begin{align*}
| \{ M : \, \exists \, \bfx \in \{0,1\}^n, M_L(\bfx) = M  \}| \leq \nchoosek{n-L+2^L}{2^L - 1}.
\end{align*}
This result implies that the maximum rate of any $L$-reconstruction code $R(n,L)$ satisfies
\begin{align}
\lim_{n \to \infty} R(n,L) \to 0 \notag
\end{align}
for $L \leq \frac{\log n}{1+\epsilon}$, and any $\epsilon>0$ (see Theorem 2.1,~\cite{CCEK17}). It remains an open problem to determine the optimal code rate for
substring lengths $L$ in the interval $[\log\,n,2\log\,n+1]$.}

Our main results are as follows. First, we show that there exists an $L$-reconstruction code with a single bit of redundancy provided that $L \geq 2 \log n + 2$, thereby improving the results of~\cite{CCEK17}. In addition, we describe an encoding scheme that requires two bits of redundancy for $L \geq 2 \log n + 4$. Second, we provide a code construction for the case $\log n < L < 2\log n$ that establishes that
\begin{align*}
\lim_{n \to \infty} R(n,L) = \begin{cases}
1, \text{ for } L \geq \lceil (1+\epsilon) \log n \rceil,\\
0, \text{ otherwise, }
\end{cases}
\end{align*}
where $0 < \epsilon < 1$. Our encoding methods rely on a novel approach termed \emph{repeat replacement}, which is of independent interest in many other string editing and design applications.

We also consider the case of noisy substring multispectra. Unlike~\cite{CCEK17}, we focus on a subclass of coverage errors expected to arise in high-accuracy sequencing platforms, and refer to the errors as \emph{bounded gap coverage errors}. More precisely, for any string $\bfx,$ one is given only a subset of the substring multispectrum, $\widehat{M}_L(\bfx) \subseteq M_L(\bfx)$. We say that the multiset $\widehat{M}_L(\bfx)$ has \emph{maximal coverage gap} $C$ if for all $i \in [n-L+1]$, there exists an $j$ such that $1 \leq |j-i| \leq C$ and $\bfx_{j,L} \in \widehat{M}_L(\bfx)$. We show that when $C$ is a constant, there exists a code of rate one with an efficient encoding/decoding procedure that allows for unique reconstruction with $L =O(\log n)$. In addition, we also present code constructions for the case when, in addition to bounded gap coverage errors, the multispectrum also contains substitution errors.

\section{Reconstruction Codebooks for $L > 2  \log n$}\label{sec:g2logn}

Let $S_L(\bfx)$ denote the {\em{set}} of all $L$-substrings of $\bfx$. If $|S_L(\bfx)| = n-L+1$, then $S_L(\bfx) = M_L(\bfx)$ and we say that $\bfx$ is $L$-substring unique. {An example of $L$-substring unique strings are (cyclic) de Bruijn strings~\cite{db} which have the property that all substrings of length $L$ appear exactly once. The number of de Bruijn strings of length $n$ and unique substring length $L$ equals $\frac{(L!)^{L^{n-1}}}{L^n}$.}

Next, recall that we refer to a codebook $\cC$ as an {\em{$L$-reconstruction code}} if for any $\bfx \in \cC$, we can recover $\bfx$ given its $L$-multispectrum $M_L(\bfx)$. The following proposition establishes simple sufficient conditions for the existence of $L$-reconstruction codes and was first reported in~\cite{ukkonen1992}.

\begin{proposition}\label{cl:uniquesubs} Suppose that $\bfx$ is $(L-1)$-substring unique. Then, $\bfx$ can be reconstructed from $S_L(\bfx) = M_L(\bfx)$. \end{proposition}
\begin{IEEEproof} For any two distinct $L$-substrings $\bfx_{i_1,L}=(x_{i_1}, x_{i_1+1}, \ldots, x_{i_1+L-1})$ and $\bfx_{i_2,L}=(x_{i_2}, x_{i_2+1}, \ldots, x_{i_2+L-1}) \in M_L(\bfx),$ the last $L-1$ bits in $\bfx_{i_1,L}$ equal to the first $L-1$ bits of $\bfx_{i_2,L}$ if and only if $(x_{i_1+1}, x_{i_1+2}, \ldots, x_{i_1+L-1}, x_{i_2,L}) \in M_L(\bfx)$. This allows one to uniquely concatenate the $L$-substrings with overlapping length $L-1$ suffix-prefixes, as is standardly done in de Bruijn graph based string assembly~\cite{Pevzner2011}.
\end{IEEEproof}

As a consequence of Proposition~\ref{cl:uniquesubs}, one straightforward approach to devising $L$-reconstruction codes is to form strings that do not have repeated substrings of length $L-1$. As a consequence, one can define an $L$-reconstruction code according to:
\begin{align*}
\cU(n,L) = \Big \{ \bfx \in \{0,1\}^n : \forall i,j, \; i \neq j, \; \bfx_{i,L-1} \neq \bfx_{j,L-1} \Big \}.
\end{align*}
Using counting arguments outlined in the Appendix, we arrive the following lower and upper bounds on $\cU(n,L)$. 

\begin{lemma}\label{lem:Unb} For $n \geq 2$, one has 
\begin{align}
2^n \cdot \left(1-\frac{ (n-L+1)^2 }{2^{L}}\right) \leq \;  &|\, \cU(n,L) \,| \leq \, 2^n \cdot  \exp \left( - \frac{n-L+1}{2^L} \cdot \left(\frac{n}{L-1} - 2 \right) \right ).
\end{align} 
\end{lemma}

For $L-1 \geq 2 \log n+1$, we also have the following corollary of Proposition~\ref{cl:uniquesubs} and Lemma~\ref{lem:Unb}.

\begin{corollary}\label{cor:red} For $L \geq 2 \log n + 2$, there exists an $L$-reconstruction code with at most one bit of redundancy. \end{corollary}
\begin{IEEEproof} To prove the result, we need to show that $\log |\,\cU(n,L)\,| \geq n - 1$ for $L \geq 2 \log n + 2$, so that the result then follows from Proposition~\ref{cl:uniquesubs}. Since $(n-L+1)^2/2^{L}$ is monotonically decreasing with $L$ and for $n\geq 2$, we have
\begin{equation*}
\log  \left(1-\frac{ (n-L+1)^2 }{2^{L}} \right) \geq \log \left(1-\frac{ (n-2 \log n -1)^2 }{4n^2} \right) 
\geq \log \left(\frac{3}{4} -\frac{1}{2n}+ \frac{4\log^2\,n+4(n-1)\log n+1}{4n^2}\right) > -1.
\end{equation*}
From Lemma~\ref{lem:Unb} it follows that $\log |\,\cU(n,L)\,| \geq n-1,$ which proves the claim.
\end{IEEEproof}

Next, we turn to the problem of designing an efficient encoder for an $L$-reconstruction code. Our constructive approach is inspired by techniques described in~\cite{Sch17} and~\cite{WiImm10} for removing \emph{runs of $0$s exceeding a certain length} from arbitrary strings. Unlike the known runlength replacement strategy, our approach -- repeat replacement -- is iterative and it may lead to the creation of new repeats in already processed substrings. 

The differences between repeat and runlength replacements are illustrated by the following two examples.

\begin{example} We first describe how runlength replacement works for the case that one wants to limit the 
length of the longest run of $0$s in a binary information string. The described approach is valid whenever the longest run is bounded from above by $\log(n+2) + 1,$ where $n$ represents the length of the information string. A detailed description  of this method may be found in~\cite{Sch17}. As an example, let $n=14$, and set the length of the 
longest allowed run of $0$s to $5$. The idea behind the approach is to delete all-zeros substrings that have length $6$ (or, in general one plus the length of the longest allowed runlength) and then append the (encoded) location of the deleted substring to the end of the original information string. 

Suppose that the length-$14$ information string of interest is
$$(0,0,0,0,0,0,0,1,0,1,0,0,0,1).$$
In the initialization step, we append the substring $(1,0)$ to the information string. For the running example, this results in a string of length $16$, namely
$$ (0,0,0,0,0,0,0,1,0,1,0,0,0,1,1,0).$$
The suffix $(1,0)$ is chosen to ensure that the string terminates with a $0$ and that 
no existing run of $0$s is extended. Each time a length-$6$ substring is deleted, $6$ symbols are appended to 
the end of the encoded string so that the length of the encoded string remains $16$. 
The first $6$ symbols of the string are $0$s. We delete this zero-substring, thereby decreasing the length of the 
first run of $0$s. We subsequently append the location of the deleted substring to the end of the string to 
make sure that during the decoding stage we can undo the deletion and recover the original information string. In our example, since the substring starts at position one, we append $(0,0,0,1)$ to the modified string to arrive at 
$$(0,1,0,1,0,0,0,1,1,0,0,0,0,1).$$
Next, we append $(1,1)$ to the end of the string to obtain a string of length $16$,
$$ (0,1,0,1,0,0,0,1,1,0,0,0,0,1,1,1).$$
The string $(1,1)$ is used to indicate that the substring $(0,0,0,1)$, which immediately precedes $(1,1)$, describes the location of a deleted all-zeros substring of length six. Since the above string no longer contains runs of $0$s of lengths greater than five, the encoding process terminates. Otherwise, one would repeat the same procedure of deleting an all-zeros substring of length $6$ and then appending $4$ bits of encoded positional information followed by the bit-pair $(1,1)$. 

The original information string can be recovered by first checking whether the last bit of the encoded string has value $1$ or $0$. If the last bit of the encoded string has value $0$, one recovers the original string by simply deleting the last two bits, as $(1,0)$ is appended by default at the beginning of the encoding procedure during the initialization step. In the example, since the last bit is a $1$, we delete the last $6$ bits, and infer that an all-zeros substring of length $6$ was deleted at position one. Re-inserting the length six all-zeros substring leads to 
$$(0,0,0,0,0,0,0,1,0,1,0,0,0,1,1,0).$$
Since the string ends with $(1,0)$, we conclude that the information string was 
$$(0,0,0,0,0,0,0,1,0,1,0,0,0,1).$$ 
\end{example}

A key observation based on the previous example is that the encoder cannot create new runs of $0$s in the information string after an all-zeros substring is removed. It is also straightforward to see that no new runs of length greater than four can be introduced by appending the positional information. Furthermore, at each removal step, the same substring (the all-zeros substring) is deleted. None of these properties hold for the case of repeat replacement, as explained in the example below.

\begin{example}\label{eq:naiveseq} Let $1^k$ and $0^k$ denote runlengths of $1$s and $0$s of length $k$, respectively, and let the information string of interest be
$$(0^{8}, 1^{4}, 0^{8}, 1^{4}, 0^{4}, 1^{8}).$$
We wish to generate a string without repeated substrings of length $8$ using a sequence replacement technique. Our approach will be the same as the one described in the previous example, in so far that repeated substrings will be deleted from the information string. Similarly to the run removal technique, after a substring is deleted, additional bits 
are appended to the string so that the deleted substring can be recovered.

In our example, the first repeated substring of length eight equals ${0}^{8}$, and it begins at position $13$. Note that removing the second instance of this substring results in the string
$$ (0^{8}, {1}^{8}, {0}^{4}, {1}^{8}),$$
which results in the creation of a new repeat that was not in the original substring (i.e., the substring ${1}^{8}$ now appears twice). Hence, the repeat removal procedure cannot be completed in one pass, as previously examined positions may give rise to new repeats.
\end{example}

As will be described shortly, in order to resolve the problems observed in Example~\ref{eq:naiveseq}, we encode the input string as follows. First, we remove fixed length substrings in multiple rounds. At each round, the length of the encoded string is reduced by exactly one. When repeats are removed, substrings encoding the location of the original substring and its repeat are appended to the string. In addition, two marker bits are appended to the encoded string to ensure accurate reconstruction. The procedure terminates when the obtained string is $(L-1)$-substring unique. To this string, we append sufficiently many $0$s to ensure that the length of the codeword is fixed -- appending $0$s does not lead to a violation of the unique $L$-substring reconstruction property as long as two marker bits are inserted into the string. Proposition~\ref{cl:uniquesubs2} rigorously formalizes the above described procedure. 

In what follows, we use $\ell(\bfv)$ to denote the length of a binary string $\bfv \in \{0,1\}^{*}$.

\begin{proposition}\label{cl:uniquesubs2} Suppose that $\bfx = ({\bfx}', {\bf0}) \in \{0,1\}^n,$ where $x_{L-1} = 1$, $\ell(\bfx') \geq L-1$, $\bfx'$ ends with a $1$, and $\bfx'$ is $(L-1)$-substring unique. Then, $\bfx$ can be uniquely reconstructed from $M_L(\bfx)$. \end{proposition}

\begin{IEEEproof} To reconstruct $\bfx$, we start by identifying the first $L-1$ positions of $\bfx$. Since $\bfx'$ has length $\ell(\bfx') \geq L-1$, and since $\bfx'$ is $(L-1)$-substring unique, the first $L-1$ bits in $\bfx$ appear as a substring in $M_L(\bfx')$ only once. Hence, $\bfx_{1,L-1} \neq \bfx_{i,L-1}$ for $i \in [\ell(\bfx')-L+2]$ (unless $i=1$). Furthermore, given that $x_{L-1}=1$, one also has $\bfx_{1,L-1} \neq \bfx_{i,L-1}$ for $i \in \{\ell(\bfx')-L+3, \ell(\bfx')-L+4, \ldots, n-L+2\}$, since the strings $\bfx_{i,L-1}$ for this range of values of $i$ end with a $0$. Therefore, $\bfx_{1,L}$ may be found by examining the prefixes of each of the $O(n)$ substrings in the spectrum. Upon identification, the substring $\bfx_{1,L}$ is removed from $M_L(\bfx)$. 

Next, we proceed with classical suffix-prefix matchings of substrings: We identify substrings in $M_L(\bfx)$ whose first $L-1$ bits match the last $L-1$ bits of $\bfx_{1,L}$.  If there is only one such substring, without counting its multiplicity, we declare it to be $\bfx_{2,L}$. If there are at least two different strings in $M_L(\bfx)$ that satisfy the matching constraint, $\bfx_{2,L}^{(1)}, \bfx_{2,L}^{(2)}$, then $\bfx_{2,L}^{(1)}$ and $\bfx_{2,L}^{(2)}$ restricted to the first $L-1$ positions equal $\bfx_{2,L-1}$. This is only possible if $\bfx_{2,L}^{(1)}$ ends with a $1$ and $\bfx_{2,L}^{(2)}$ ends with a $0$, or vice versa. Since $\bfx'$ is $(L-1)$-substring unique and ${\bfx} = ({\bfx'}, {\bf0})$, one of the two substrings $\bfx_{2,L}^{(1)}, \bfx_{2,L}^{(2)}$ has to appear in $\bfx$ starting at position $j,$ where $j \geq \ell(\bfx')-L+2$. Since by assumption $\bfx_{2,L}^{(2)}$ ends with a $0$ and due to the fact that $\bfx$ ends with $n-\ell(\bfx')$ $0$s, it follows that $\bfx_{2,L}=\bfx_{2,L}^{(1)}$. We remove one instance of the substring $\bfx_{2,L}^{(1)}$ from the multi-set $M_L(\bfx)$ and proceed. We continue in the same manner until the multiset $M_L(\bfx)$ contains only all-zeros substrings. Then, since $\bfx$ is of the form $\bfx=(\bfx', \bf0)$, we set the remaining bits of $\bfx$ to $0$ and complete the decoding.\end{IEEEproof} 

In order to describe our procedure for generating $(L-1)$-substring unique strings, we introduce the notion of \textit{nonoverlapping repeated substrings} and \textit{overlapping repeated substrings}. We say that a string $\bfx \in \{0,1\}^n$ has a nonoverlapping repeated substring of length $L-1$ at the positions $(i,j), i < j,$ if $\bfx_{i,L-1} = \bfx_{j,L-1}$ and $j-i \geq L-1$. {For example, the string $(1,0,1)$ is a non-overlapping repeated substring of $\bfx=(1,0,1,0,0,1,0,1)$ at the positions $(1,6)$.} On the other hand, if $i<j$ and $\bfx_{i,L-1} =\bfx_{j,L-1}$ for $j-i < L-1$, we say that $\bfx$ has an overlapping repeated substring of length $L-1$ at the positions $(i,j)$. {For example, the string $(1,0,1)$ is an overlapping repeated substring of $\bfx=(1,0,1,0,1,0,0)$ at the positions $(1,3)$.}
In either of the two cases, we say that $\bfx$ has repeats at positions $(i,j)$, $i<j$, of length $L-1$. 

For an integer $k \in [n]$, let $B(k) \in \{0,1\}^{ \log n }$ denote its binary representation of length $\log n$, and recall that the length of a string $\bfv \in \{0,1\}^{*}$ is denoted by $\ell(\bfv)$. Unless stated otherwise, we assume throughout the remainder of this section that $L=  2  \log n  + 4$. 

\begin{proposition}\label{cl:repeatp} Suppose that the string $\bfx \in \{0,1\}^n$ has an overlapping repeated substring of length $L-1$ at the positions $(i,j)$. Then, $p(\bfx_{i, L-1+j-i}) {\leq } j-i$.  \end{proposition}
\begin{IEEEproof} If $\bfx_{i,L-1} = \bfx_{j,L-1}$, then $(x_j, x_{j+1}, \ldots, x_{j+L-2}) = (x_i, x_{i+1}, \ldots, x_{i+L-2})$ and in particular, $(x_j, x_{j+1}, \ldots, x_{i+L-2})$ $=(x_i, x_{i+1}, \ldots, x_{2i - j + L -2})$. Thus, $(x_{i+p}, x_{i+p+1}, \ldots, x_{j+p-1})=$ $(x_{j+p}, x_{j+p+1}, \ldots, x_{2j+p-i-1})$, for $0 \leq p \leq (L-1)-(j-i)$, which implies that $p(\bfx_{i, L-1+j-i})=j-i$.
\end{IEEEproof}

We are now ready to describe the repeat replacement encoder $\cE_{RR}$, the steps of which are outlined in Algorithm~1. The input of the encoder is a string $\bfx_{I} \in \{0,1\}^{n}$ such that its $(L-1)$-st and last bit are both equal to $1$. The output of $\cE_{RR}$ is a string $\bfx \in \{0,1\}^{*},$ such that $\ell(\bfx) \leq \ell(\bfx_I)$, $\bfx$ is $(L-1)$-substring unique, and the $(L-1)$-st and last bit of $\bfx$ are both equal to $1$.

\begin{algorithm}[H]
\caption{Repeat replacement encoder $\cE_{RR}$ for generating $(L-1)$-substring unique strings.}
\begin{algorithmic}[1]
\STATE If $\bfx_I$ is $(L-1)$-substring unique, set $\bfx = \bfx_I$ and STOP. Otherwise, set $\bfx^{(0)} = \bfx_I$, and 
let $k=1$.
  \STATE Suppose that $\bfx^{(k-1)}$ has a repeat at positions $(i,j)$ of length $L-1$. Let $\bfx^{(k)}$ be obtained by 
  deleting $\bfx^{(k-1)}_{j,L-1}$ from $\bfx^{(k-1)}$ and subsequently appending the string $(B(i), B(j),0,1)$ at the end of the generated string.
  \STATE If $x^{(k)}_{L-1}=0$, i.e., if the $(L-1)\,$-st bit of $\bfx^{(k)}$ equals to $0$, reset $x^{(k)}_{L-1}=1$ and update the 
  last two bits of $\bfx^{(k)}$ to $(1,1)$. If $\bfx^{(k)}$ is $(L-1)$-substring unique, set $\bfx= \bfx^{(k)}$ and STOP. Otherwise, set $k=k+1$, and go to Step 2.
\end{algorithmic}
\end{algorithm}

{Note that if the string does not contain repeats, its length clearly remains the same as the algorithm terminates immediately.} Furthermore, in Step 2, for each possible value of $k$, we have $\ell(\bfx^{(k-1)}_{j,L-1}) = L-1$ and $\ell(B(i), B(j),0,1) = 2\log n + 2= L-2$; as a result, it holds that $\ell(\bfx^{(k)}) = \ell(\bfx^{(k-1)})-1$. Furthermore, the following claim holds true.

\begin{proposition}\label{cl:propsrr} Suppose that $\bfx = \cE_{RR}(\bfx_I)$. Then, $\bfx$ is $(L-1)$-substring unique, 
with its $(L-1)$-st and last bit equal to $1$, and $\ell(\bfx) \geq L-1$.  \end{proposition}
\begin{IEEEproof} We first show that $\ell(\bfx) \geq L-1$. Suppose on the contrary that $\ell(\bfx)=N< L-1$. Then, during the encoding procedure we had to encounter $\ell(\bfx^{(k-1)})=N+1$ and  $\ell(\bfx^{(k)})=N$ for some $k\geq1$. Since $\ell(\bfx^{(k)}) = \ell(\bfx^{(k-1)}) - 1$, it follows that $N=L-1,$ as any string of length $L-1$ is $(L-1)$-substring unique, {and the algorithm terminates at Step 3}. In this case, we also have $\ell(\bfx)=L-1,$ {which contradicts the assumption that $\ell(\bfx)=N< L-1$.} {Note that in this case} $\bfx$ ends with a $1$ since $(B(i), B(j),0,1)$ ends with a $1$.

Suppose now that $\ell(\bfx)> L-1$. Then, the algorithm terminated either at Step 1 or Step 3, with $\bfx = \bfx^{(k)}$ for some $k \geq 0$. Since the algorithm terminated, $\bfx=\bfx^{(k)}$ had to be $(L-1)$-substring unique.

The fact that $\bfx$ ends with a $1$ and that the $(L-1)$-st bit of $\bfx$ equals $1$ follows immediately from the description of the encoding process. In particular, if $\bfx = \bfx_I$, given that the input $\bfx_I$ has a $1$ at the last and $(L-1)$-st position, the same has to be true for $\bfx$. Otherwise, if $\bfx \neq \bfx_I$, the result follows since $(B(i), B(j),0,1)$ terminates with a $1$ and in Step 3, if necessary, the $(L-1)$-st bit of $\bfx$ is set to $1$.
\end{IEEEproof}

In what follows, we describe the decoding procedure that allows us to recover $\bfx_I$ from $\bfx$.

\begin{lemma}\label{lem:RREenc} Suppose that $\bfx = \cE_{RR}(\bfx_I)$. Given $\bfx$, one can uniquely 
recover $\bfx_I$. 
\end{lemma}
\begin{IEEEproof} Suppose that the encoding process terminates after $t$ rounds. If $t=0$, the result follows since $\ell(\bfx)=n$, which establishes that $\bfx_I = \bfx$. 

Now, assume that $t>0$, so that $\ell(\bfx)<n$, which in turn implies $\bfx \neq \bfx_I$. In what follows, we show that it is possible to uniquely recover $\bfx^{(k-1)}$ from $\bfx^{(k)}$. Clearly, $\ell(\bfx) = n - t$. 

We start by removing the last $L-2$ bits of $\bfx$ which encode 
$(B(i), B(j),b,1)$, where $b \in \{{0,1\}}$. The resulting string is denoted by $\bfx'$. 
If $b=1$, we set the $(L-1)$-th bit of $\bfx'$ to $0$. If $j-i \geq L-1$, then we reinsert a nonoverlapping 
repeated substring into $\bfx'$ to obtain $\bfx^{(k-1)}$. More precisely, we insert a substring of length $L-1$ that starts at position $i$ in $\bfx'$ into the same string, but at position $j$. Otherwise, if $j-i < L-1$, we identify a substring of length $L-1$ with period $j-i$ whose first $j-i$ positions match those in $\bfx'$ starting at position $i$ and ending at position $j-1$. 
We insert the identified substring into position $j$ of $\bfx'$ to arrive at $\bfx^{(k-1)}$ from $\bfx^{(k)}$ according to Proposition~\ref{cl:repeatp}. This procedure is repeated iteratively and recovers $\bfx_I$ exactly.
\end{IEEEproof}

The following algorithm shows how to leverage the repeat replacement encoder $\cE_{RR}$ for the purpose of unique $L$-reconstruction encoding. The $L$-reconstruction code encoder $\cE_{LR}$ described in Algorithm~2 takes as its input a binary string $\bfx_I$ of length $n-2$ and the outputs a binary string $\bfx$ of length $n$.

\begin{algorithm}[H]
\caption{Encoder $\cE_{LR}$ for an $L$-reconstruction code.}
\begin{algorithmic}[1]
  \STATE Let $\bfx_I \in \{0,1\}^{n-2}$. If the $L-1\,$th bit of $\bfx_I$ is $1$, append $(0,1)$ to the string. 
  Otherwise, set the $L-1\,$th bit of $\bfx_I$ to $1$, and append $(1,1)$ to the string.
  \STATE If $\ell(\cE_{RR}(\bfx_I)) = n$, set $\bfx = \cE_{RR}(\bfx_I)$. Otherwise, append to $\cE_{RR}(\bfx_I)$ as many 
  $0$s as needed to make the string $\bfx$ have length $n$.
\end{algorithmic}
\end{algorithm}

The next result follows from the description of $\cE_{LR}$, Propocition~\ref{cl:uniquesubs2}, and Lemma~\ref{lem:RREenc}.

\begin{theorem}\label{th:recU} There exists an $L$-reconstruction code for all values of $L \geq 2\log n + 4$ that has cardinality $2^{n-2}$. \end{theorem}
\begin{IEEEproof} We outline the proof for the case $L=2\log n + 4$. A simple modification of the encoder $\cE_{RR}$, described in the Appendix, shows that the same result is true for values of $L$  exceeding $2 \log n + 4$.

First, it follows from the description of the encoder $\cE_{LR}$ that any codeword can be uniquely reconstructed given its $L$-multispectrum from Proposition~\ref{cl:uniquesubs2} and Proposition~\ref{cl:propsrr}, 
since $\ell(\cE_{RR}(\bfx_I)) \geq L-1$, $\cE_{RR}(\bfx_I)$ is $(L-1)$-substring unique, the $(L-1)\,$th bit of 
$\cE_{RR}(\bfx_I)$ has value $1$, and $\cE_{RR}(\bfx_I)$ ends with a $1$. 
We only need to show that given any codeword $\bfx$, it is possible to uniquely recover $\bfx_I$, which is the input to the encoder $\cE_{LR}$ in Algorithm~2. Let $\bfz' =\cE_{RR}(\bfx')$, where $\bfx'$ is the input of Step 2 of the encoder $\cE_{RR}$, so that $\bfx = (\bfz', \bf0)$. Recall from Algorithm~1 that $\bfz'$ ends with the symbol $1$ so that given any codeword $\bfx$, it is possible to extract the vector $\bfz'$. From Lemma~\ref{lem:RREenc}, we can then recover $\bfx'$ from $\bfz'$. 
If the last two bits of $\bfx'$ are $(0,1)$, then $\bfx_I \in \{0,1\}^{n-2}$ is equal to the first $n-2$ bits of $\bfx'$. 
Otherwise, $\bfx_I$ is equal to the first $n-2$ bits of $\bfx'$ after the $(L-1)\,$th bit of $\bfx'$ is set to $0$. 
\end{IEEEproof}

\section{The case $\log n < L < 2 \log n$}\label{sec:l2logn}

We now turn our attention to the case $L < 2 \log n$. In the previous section, we showed in Corollary~\ref{cor:red} that for $L \geq 2 \log n + 2$, one has $\lim_{n \to \infty} R(n,L) \to 1$. In what follows, we describe an $L$-reconstruction code for $L=\log n + 2\log \log n + 8$ whose rate approaches $1$ as $n \to \infty$. In Theorem~\ref{th:SM}, we will use this $L$-reconstruction code to show that $\lim_{n \to \infty} R(n, \lceil (1+\epsilon) \log n \rceil) \to 1$ for any constant $\epsilon$ such that $0 < \epsilon < 1$.

We start by introducing some relevant notation. Let $\bfx \in \{0,1\}^n$ and suppose that $\bfx$ has a repeat at positions $(i,j)$ and another not necessarily distinct repeat at positions $(i', j')$, both of length $L-1$. We write $(i,j) \leq (i',j')$ if $j \leq j'$, and similarly $(i,j) < (i',j')$ if $j < j'$. If $\bfx$ is not $(L-1)$-substring unique, we say that the repeat at $(i,j)$ in $\bfx$ is
\textit{primal} if for any repeat at $(i',j')$ in $\bfx$, $(i,j) \leq (i',j')$. Let $B_r : [n] \to \{0,1\}^{\log n + 1}$ be an injective mapping such that for an integer $i \in [n]$, $B_r(i) \in \{0,1\}^{\log n + 1}$ represents a string that does not contain a run of $0$s of length $\geq 2 \log \log n$. It will be verified in Proposition~\ref{cl:labeling} that such a labeling is possible, since there are at least $n$ strings in $\{0,1\}^{\log n + 1}$ that have no runs of $0$s of lengths exceeding $2 \log \log n-1$.

We now outline the main ideas of the encoding and decoding procedure. The goal of the encoder is to produce a string $\bfx$ which is $(L-1)$-substring unique. At a high level, our approach is similar to the approach described in the previous section. In particular, we rely on a repeat encoder, which we refer to as the primal repeat replacement encoder, that removes repeats of substrings of length $L-1$ and, in the process of doing so, compresses the input string. As in the previous section, our codewords are obtained by appending $0$s to the output of the primal repeat replacement encoder, so that according to Proposition~\ref{cl:uniquesubs2}, the resulting string is uniquely reconstructable based on its $L$-multispectrum. 

There are two main differences between the primal repeat replacement encoder and the repeat replacement encoder from the previous section: 1) The input to the primal repeat replacement encoder is required to be free of any runs of $0$s of length $\geq 2 \log \log n$; 2) Upon removal of repeated substrings, the primal repeat replacement encoder introduces \emph{marker} runs of $0$s of length $2\log \log n$ at the position of the repeated substring. As a result, the only positional information that needs to be included into the string to make the encoding procedure invertible is the location of the first occurrence of the repeated substring -- the position of the second repeat can be inferred from the runlength of $0$s and the runlength encoded location of the first substring. In comparison, the generic repeat encoder from the previous section has to include information about the positions of \emph{both} occurrences of the repeated substring within the string to be encoded.

We now turn our attention to the primal repeat replacement encoder. We start with an information string $\bfx_I \in \{0,1\}^{n}$ that does not contain runlengths of $0$s of length $\geq 2\log \log n$, ends with a $1$ and has a $1$ at position $L-1$. We remove repeats from $\bfx_I$ one at a time so that at the end of the process, the encoded string is $(L-1)$-substring unique. At each step, we choose to remove the primal repeat. Suppose, for instance, that $\bfx_I$ has a primal repeat at $(i,j)$. We first remove the substring $\bfx_{I \, j,L-1}$ from $\bfx_I$. We then insert at position $j$ the substring $(0,0,\ldots, 0, 1)$ of length $2\log \log n + 1$. The substring $(0,0,\ldots, 0, 1)$ is a marker that indicates to the decoder the position of a removed repeated substring. Immediately following $(0,0,\ldots, 0, 1)$, we insert $(B_r(i),1) \in \{0,1\}^{\log n+2}$, encoding the value of the first position $i$ so that is satisfies a runlength constraint, with an added bit $1$. If needed, we also update the resulting string so that it has value $1$ in position $L-1$, and in the process of doing so, we append three bits to the end of the string. Notice that according to this process, we have removed a substring from $\bfx_I$ of length $L-1$ and inserted $\log n + 2\log \log n + 6 = L-2$ bits so that the length of the resulting string is $n-1$. This process is repeated until the resulting encoded string does not contain any repeated substrings of length $L-1$. Therefore, similar to what we established for $\cE_{RR}$, the encoding procedure terminates after at most $n-L+1$ repeated substrings are removed and the resulting encoded string has length $\geq L-1$.

The decoder operates as follows. Let $\bfx$ denote the output of the primal repeat replacement encoder. If $\ell(\bfx) = n$, then we conclude that $\bfx_I = \bfx$. Otherwise, we begin by removing the last three bits in $\bfx$. If this substring equals $(1,1,1)$, we set the $(L-1)$-th bit of $\bfx$ to $0$. Otherwise, we perform no updates. 
We then locate the last occurrence of the substring $(0,0,\ldots,0,1)$ of length $2\log \log n + 1$ in $\bfx$. The $\log n+1$ bits immediately following $(0,0,\ldots,0,1)$ correspond to $B_r(i)$, indicating that the substring $\bfz_{i,L-1}$ was removed at position $j$ during encoding. Proposition~\ref{cl:invert} establishes that the removal of this repeat can be inverted by replacing the substring $(0,0,\ldots,0,1,B_r(i),1)$ with the substring at position $i$ in $\bfx$. This process is continued until the resulting string has length $n$ and the vector $\bfx_I$ is recovered.

Next, we describe the encoding and decoding processes in detail and show that the rate of the encoder approaches one for sufficiently large $n$. We start with a rigorous outline of the primal repeat replacement encoder in Algorithm~3 (Recall that $\bfx_I \in \{0,1\}^n$ has no runs of $0$s of length $\geq 2\log \log n$, and that the (L-1)-st and last bit of $\bfx_{I}$ are equal to $1$).

\begin{algorithm}[H]
\caption{Primal repeat replacement encoder $\cE_{Prr}$ for generating $(L-1)$-substring unique strings.}
\begin{algorithmic}[1]
\STATE If $\bfx_I$ is $(L-1)$-substring unique, set $\bfx = \bfx_I$ and STOP. Otherwise, set $\bfx^{(0)} = \bfx_I$, and let $k=1$.
  \STATE Suppose that $\bfx^{(k-1)}$ has a repeat at $(i,j)$ of length $L-1,$ where $(i,j)$ is primal. Generate $\bfx^{(k)}$ by replacing $\bfx^{(k-1)}_{j,L-1}$ with the string $(0,0,\ldots,0,1,B_r(i),1)$ of length $L-5$.
  \STATE If $x_{L-1}^{(k)} = 0$, update $x_{L-1}^{(k)} = 1$ and append $(1,1,1)$ to $\bfx^{(k)}$. Otherwise, append $(1,0,1)$ to $\bfx^{(k)}$. If $\bfx^{(k)}$ is $(L-1)$-substring unique, set $\bfx = \bfx^{(k)}$ and STOP. Otherwise, set $k=k+1$, and go to Step 2.
\end{algorithmic}
\end{algorithm}

The next proposition is an analogue of Proposition~\ref{cl:propsrr} for primal replacement encoding, and it may be proved using the same arguments.

\begin{proposition} Suppose that $\bfx = \cE_{Prr}(\bfx_I)$. Then, $\bfx$ is $(L-1)$-substring unique, has length $\geq L-1$, and takes the value $1$ at the last and $(L-1)$-th position. \end{proposition}

The following proposition establishes that one can recover $\bfx_I$ from $\cE_{Prr}(\bfx_I)$. 

\begin{proposition}\label{cl:invert}  For $\cE_{Prr}$ and $k \geq 1$, the string $\bfx^{(k-1)}$ can be reconstructed from $\bfx^{(k)}$.    \end{proposition}
\begin{IEEEproof} We start by removing the last three bits from $\bfx^{(k)}$. If these last three bits are all equal to $1$, we update $\bfx^{(k)}$ so that its $(L-1)$-th bit has value $0$. Suppose that the last occurrence of the string $(0,0,\ldots,0,1)$ of length $2\log \log n + 1$ in $\bfx^{(k)}$ is at position $j$. Then, $\bfx^{(k)}_{j,L-5}=(0,0,\ldots,0,1,$ $B_r(i),1)$. Similar to the proof of Lemma~\ref{lem:RREenc}, there are two cases to consider. If $(i,j)$ represents a nonoverlapping repeated substring, then we obtain $\hat{\bfx}^{(k-1)}$ by replacing $\bfx^{(k)}_{j,L-5}$ with $\bfx^{(k)}_{i,L-1}$. Otherwise, if $(i,j)$ represents an overlapping repeated substring, so that $j-i<L-1$, we replace $\bfx^{(k)}_{j,L-5}$ with a string of length $L-1$ and period $j-i$ from $\bfx^{(k)}_{i,j-i}$.

Assume next that the string generated according to the previous procedure does not equal $\bfx^{(k-1)}$. This is only possible if $\bfx^{(k)}$ is the result of removing a repeat at $(i',j')$ from $\bfx^{(k-1)}$ such that $j' < j$. Since there is a run  of $0$s of length at least $2 \log \log n$ starting at position $j$, we know that at some point during encoding, the encoder $\cE_{Prr}$ removed a repeated substring of length $L-1$ from position $j$. Since the encoder $\cE_{Prr}$ always chooses the primal repeat to remove according to Step 2 of Algorithm~3, if $j'<j$, we would had to have a prefix of $\bfx^{(k-1)}_{j,L-1}$ removed and replaced with a suffix of $(0,0,\ldots,0,1,B_r(i'),1)$. Since any proper suffix of $(0,0,\ldots,0,1,B_r(i'),1)$ does not contain runs of $0$s of length $\geq 2\log \log n$, we arrive at a contradiction as $\bfx^{(k)}_{j,2\log \log n + 1} \neq (0,0,\ldots,0,1)$. \end{IEEEproof}

The next lemma follows from the two previous results and can be proved using the same ideas as those described in Lemma~\ref{lem:RREenc}.

\begin{lemma} Suppose that $\bfx = \cE_{Prr}(\bfx_I)$. Given $\bfx$, one can recover $\bfx_I$. \end{lemma}

The encoder $\cE_{Plr}$ of the $L$-reconstruction code is the same as the encoder $\cE_{LR}$, except for the fact that $\cE_{Plr}$ calls $\cE_{Prr}$ rather than $\cE_{RR}$, and that the input $\bfx_I$ is required to satisfy a runlength constraint. The input to $\cE_{Plr}$ in Algorithm~4 is $\bfx_I \in \{0,1\}^{n-3},$ where $\bfx_I$ does not contain any runs of $0$s of lengths $\geq 2 \log \log n$.

\begin{algorithm}[H]
\caption{The encoder $\cE_{Plr}$ for a primal $L$-reconstruction code.}
\begin{algorithmic}[1]
  \STATE If the $(L-1)$-st bit of $\bfx_I$ equals $1$, append $(1,0,1)$ to $\bfx_I$. Otherwise, set the value of the $(L-1)$-st bit to $1$, and append $(1,1,1)$ to the resulting string.
  \STATE If $\ell(\cE_{Prr}(\bfx_I)) = n$, set $\bfx = \cE_{Prr}(\bfx_I)$. Otherwise, append as many $0$s to $\cE_{Prr}(\bfx_I)$ as needed for $\bfx$ to have length $n$.
\end{algorithmic}
\end{algorithm}

It is straightforward to prove the following result.

\begin{lemma}\label{lem:Mlr} The string $\cE_{Plr}(\bfx_I)$ belongs to an $L$-reconstruction code. \end{lemma}

We now turn our attention to showing that the above construction results in a code of rate one. We start by 
characterizing the number of possible choices for the vector $\bfx_I \in \{0,1\}^{n-3}$ that can be used as inputs to $\cE_{Plr}$.

\begin{proposition}\label{cl:u} There are at least 
$$ \Bigg ( \frac{n-3}{4} \Big ( 1 - \frac{1}{\log (n-3)} \Big) \Bigg )^{\lfloor \frac{n-3}{\log (n-3)} \rfloor}$$
possible different input strings $\bfx_I$ for Algorithm~4.
\end{proposition}
\begin{IEEEproof} Let $n' = n-3$, and suppose that $(\bfx_I^1, \bfx_I^2, \ldots, \bfx_I^{\lfloor \frac{n'}{\log n'} \rfloor})$ constitute the first $(\log n') \cdot \lfloor \frac{n'}{\log n'} \rfloor$ bits of $\bfx_I$. For $i \in \big [\lfloor \frac{n'}{\log n'} \rfloor \big ]$, we set the first and last bits of $\bfx_I^i \in \{0,1\}^{\log n'}$ to $1$. In addition, we require that $\bfx_I^i$ not have any runs of $0$s exceeding $2 \log \log n' < 2 \log \log n$. Using similar counting arguments as those used in the proof of Lemma~\ref{lem:Unb}, one can show that the number of  different choices for $\bfx_I^i$ is at least $2^{\log n' - 2} - (\log n') \cdot 2^{\log n' -2 - 2\log \log n'} = \frac{n'}{4} \cdot \left( 1 - \frac{1}{\log n'} \right)$. Since we are to concatenate $\frac{n'}{\log n'}$ such strings, we have at least 
$$ \Bigg ( \frac{n'}{4} \Big ( 1 - \frac{1}{\log n'}  \Big ) \Bigg)^{\lfloor \frac{n'}{\log n'} \rfloor }$$
choices for the first $(\log n') \cdot \lfloor \frac{n'}{\log n'} \rfloor$ bits of $\bfx_I$. The remaining $n' - (\log n') \cdot \lfloor \frac{n'}{\log n'} \rfloor$ bits of $\bfx_I$ are set to $1$. Substituting $n'=n-3$ produces the claimed result.
\end{IEEEproof}

\begin{proposition}\label{cl:labeling} There exists an invertible encoder $B_r : [n] \to \{0,1\}^{\log n + 1}$ such that for any $i \in [n]$, the length of the longest run of $0$s in $B_r(i)$ is at most $2 \log \log n - 1$.  \end{proposition}
\begin{IEEEproof} The statement follows since we can show that there are at least $n$ strings in $\{0,1\}^{\log n + 1}$ that do not have any runs $\geq 2 \log \log n$ and use the same counting arguments as in Lemma~\ref{lem:Unb}.
\end{IEEEproof}

The above propositions and Lemma~\ref{lem:Mlr} may be used to establish the following theorem.

\begin{theorem}\label{th:SM} Let $L = (1+\epsilon) \log n,$ where $0 < \epsilon < 1$. Then, $$\lim_{n \to \infty} R(n,L) \to 1.$$ \end{theorem}
\begin{IEEEproof} According to Proposition~\ref{cl:u}, there are at least $\Bigg ( \frac{n-3}{4} \Big ( 1 - \frac{1}{\log (n-3)} \Big) \Bigg )^{\lfloor \frac{n-3}{\log (n-3)} \rfloor}$ choices for $\bfx_I$. Thus,
\begin{align*}
R(n,\log n + 2 \log \log n + 5) \geq \frac{1}{n} \cdot \left( \frac{n-3}{\log (n-3)}-1 \right) \log \left( \frac{n-3}{4}\right) + \frac{1}{n} \cdot \left( \frac{n-3}{\log (n-3)} -1 \right) \log \left(1 - \frac{1}{\log (n-3)} \right).
\end{align*}
The right hand side of the above expression converges to $1$ as $n \to \infty$. As before, the case $L > \log n + 2\log \log n + 5$ can be handled through a simple modification of the primal repeat replacement encoder, described in Appendix~\ref{app:encLSB}.
\end{IEEEproof}

\section{String Reconstruction From Noisy $L$-Multispectra}\label{sec:errors}

We consider next the problem of reconstructing strings from their noisy $L$-multispectrum. We are interested in two types of spectrum errors. The first type of errors, usually referred to as \textit{coverage errors}, occurs when substrings are removed from the $L$-multispectrum. Unlike the related work~\cite{CCEK17}, we focus on a special new type of coverage errors, termed \emph{sparse coverage errors}. These errors capture the fact that modern high throughput sequencing devices do not introduce long bursts of coverage errors. The second type of errors, known as \textit{spurious errors}, occur when individual substrings in the $L$-multispectrum are subjected to substitution errors. In our subsequent derivations, we introduce some simple constructions that extend the setup introduced in Section~\ref{sec:g2logn} so that it applies to noisy spectral reconstruction. The presented codes have code rates that converge to one with increasing block length, and may be encoded and decoded in polynomial time.

\subsection{Codes for Coverage Errors}

We start with a construction for codes capable of correcting sparse coverage errors only. As before, we use $M_L(\bfx)$ to denote the $L$-multispectrum of a binary string $\bfx$. The spectrum obtained from a noisy readout device presented with the string $\bfx$ is denoted by $\widehat{M}_L(\bfx)$. We require that $\widehat{M}_L(\bfx) \subseteq M_L(\bfx)$ - i.e., that any substring in $\widehat{M}_L(\bfx)$ also belongs to $M_L(\bfx)$. The spectrum $\widehat{M}_L(\bfx)$ is said to have \emph{maximal coverage gap} $G$ if $G$ is the maximum number of consecutive positions for which the substrings starting at those positions are not included in $\widehat{M}_L(\bfx)$. In addition, we say that a code is an $(L,G)$-reconstruction code if given $\widehat{M}_L(\bfx)$ with maximal coverage gap $G$, it is possible to uniquely reconstruct $\bfx$ whenever the string belongs to the code. The next example illustrates the notion of a coverage gap.

\begin{example}\label{ex:gapex} Let $\bfx = (0,1,1,0,1,0,0)$. Then, $M_4(\bfx) = \Big \{ (0,1,1,0), (1,1,0,1), (1,0,1,0), (0,1,0,0) \Big \}$. 
Assume that the observed noisy spectrum equals $\widehat{M}_4(\bfx) = \Big \{ (1,0,1,0), (0,1,0,0) \Big \}$. Then, $\widehat{M}_4(\bfx)$ has maximal coverage gap equal to two since there are two substrings in ${M}_4(\bfx) \setminus \widehat{M}_4(\bfx),$ and the two ``missing'' substrings start at two consecutive positions, $1$ and $2$.\end{example}

We present next a construction for an $(L,G)$-reconstruction code. The intuition behind our approach is as follows. Suppose that one is given a prefix of $\bfx$ of length $L$, denoted by $\bfx_L$. The idea is to iteratively extend ${\bfx}_L$ and thereby obtain a longer prefix of $\bfx$ given the set  $\widehat{M}_L(\bfx)$ and the side information that $\widehat{M}_L(\bfx)$ has maximal coverage gap $G$. In order to be able to perform this extension, we restrict $\bfx$ to be $\hat{L}$-substring unique, where $\hat{L} < L-G$. Under this setup, $\bfx_L$ contains at least $G+2$ substrings of length $\hat{L}$, so that $|M_{\hat{L}}(\bfx_L)| = G+2$. Since $\widehat{M}_L(\bfx)$ has maximal coverage gap $G$, there exists some $\bfx' \in \widehat{M}_L(\bfx)$ and some $j > 1$ such that $\bfx'_{1,\hat{L}}={(\bfx_{L})_{j,\hat{L}}}.$ Therefore, we can append the last $j-1$ bits of $\bfx'$ to $\bfx_L$ to generate a longer prefix of $\bfx$. Continuing in this manner, it is possible to fully recover $\bfx$ given a few additional boundary constraints.

Recall the proof of Proposition~\ref{cl:uniquesubs2} where in order to reconstruct a string $\bfx$ from its $L$-multispectrum, we relied on the assumption that the first $L-1$ bits of $\bfx$ do not appear more than once in $M_L(\bfx)$. The reasons why this condition was sufficient in the given setting are that 1) every substring of $\bfx$ of length $L$ appears exactly once in $M_L(\bfx)$ and 2) if $\bfx' \in M_L(\bfx)$ is such that $\bfx' \neq \bfx_{1,L}$, then the prefix of length $L-1$ of $\bfx'$ necessarily appears as a suffix of some other substring in $M_L(\bfx)$. For the case where $\widehat{M}_L(\bfx)$ has maximal coverage gap $G$, the substrings $\bfx_{1,L}, \bfx_{2,L}, \ldots, \bfx_{G,L}$ may not be in $\widehat{M}_L(\bfx)$, although at least one of the strings $\bfx_{1,L}, \bfx_{2,L}, \ldots, \bfx_{G,L}, \bfx_{G+1,L}$ must belong to $\widehat{M}_L(\bfx)$. Similarly, the substrings $\bfx_{n-L-G+2,L}, \bfx_{n-L-G+3,L}, \ldots, \bfx_{n-L+1,L}$ may not belong to $\widehat{M}_L(\bfx)$, while an extended subset has to include at least one substring that lies in $\widehat{M}_L(\bfx)$. To account for these differences, we need to modify the encoders from Section~\ref{sec:g2logn}, which we do in what follows. 

We describe two encoders that are used to generate the codewords of an $(L,G)$-reconstruction code $\cC_G(n)$. 
First, we discuss what we refer to as the \emph{gap repeat replacement encoder}. The gap repeat replacement encoder $\cE_{Grr}$ functions similarly to the repeat replacement encoder described in Section~\ref{sec:g2logn}. The main difference between the gap repeat replacement encoder and the repeat replacement encoder is that the former ensures that the first $G+1$ substrings of length $\hat{L}$ that appear in the encoded string end with a $1$, whereas the repeat replacement encoder only ensures that the first substring of length $L-1$ in the encoded string ends with a $1$. 

Let $\hat{L} = 2\log n + 3 + G$ and suppose that $L = \hat{L} + G + 1 = 2 \log n + 2G + 4$. Recall that the parameter $G \geq 1$ is an integer which represents the coverage gap. The input to $\cE_{Grr}$ is a string $\bfx_I \in \{0,1\}^n$ that takes the value $1$ at all positions in $\{ \hat{L},\hat{L}+1, \ldots,\hat{L}+G \}$. Furthermore, the last bit of $\bfx_I$ also equals $1$. The output of $\cE_{Grr}$ is a string $\bfx \in \{0,1\}^{*}$ such that $\bfx$ is $\hat{L}$-substring unique, $\ell(\bfx) \leq \ell(\bfx_I)$, and the last bit of $\bfx$ has value $1$. In addition, if $\ell(\bfx) \geq \hat{L}+G$, then $\bfx_{ \hat{L},G+1} = (1,1,\ldots,1)$. Otherwise, if $\ell(\bfx) < \hat{L}+G$, then $\bfx_{ \hat{L},\ell(\bfx)- \hat{L} +1} = (1,1,\ldots,1)$.

\begin{algorithm}[H]
\caption{The gap repeat replacement encoder $\cE_{Grr}$ for generating $\hat{L}$-substring unique strings.}
\begin{algorithmic}[1]
\STATE If $\bfx_I$ is $\hat{L}$-substring unique, set $\bfx = \bfx_I$ and STOP. Otherwise, set $\bfx^{(0)} = \bfx_I$, and 
let $k=1$.
  \STATE Suppose that $\bfx^{(k-1)}$ has a repeat at positions $(i,j)$ of length $\hat{L}$. Let $\bfx^{(k)}$ be obtained by 
  deleting $\bfx^{(k-1)}_{j,\hat{L}}$ from $\bfx^{(k-1)}$ and subsequently appending the string $(B(i), B(j),1,1,\ldots,1,1) \in \{0,1\}^{\hat{L}-1}$ to the end of the generated string.
  \STATE If $\ell(\bfx) \geq \hat{L}+G$ and ${\bfx}^{(k)}_{\hat{L},G+1}\neq (1,1,\ldots,1)$, set the last $G + 2$ bits of ${\bfx}^{(k)}$ to $({\bfx}^{(k)}_{\hat{L},G+1}, 1)$ and update ${\bfx}^{(k)}_{\hat{L},G+1} = (1,1,\ldots,1)$. If $\bfx^{(k)}$ is $\hat{L}$-substring unique, set $\bfx= \bfx^{(k)}$ and STOP. Otherwise, set $k=k+1$, and go to Step 2.
\end{algorithmic}
\end{algorithm}

The following proposition easily follows from the techniques described in the previous section.

\begin{proposition}\label{cl:Grr} The string $\bfx$ is $\hat{L}$-substring unique, $\hat{L} \leq \ell(\bfx) \leq \ell(\bfx_I)$, and the last bit of $\bfx$ has value $1$. In addition, if $\ell(\bfx) \geq \hat{L}+G$, then $\bfx_{ \hat{L},G+1} = (1,1,\ldots,1)$. Otherwise, if $\ell(\bfx) < \hat{L}+G$, then $\bfx_{ \hat{L},\ell(\bfx)- \hat{L} +1} = (1,1,\ldots,1)$.  \end{proposition}
\begin{IEEEproof} The fact that $\bfx$ is $\hat{L}$-substring unique follows from the same proof approach outlined in Proposition~\ref{cl:propsrr}. For $\ell(\bfx) \geq \hat{L}+G$, it follows that $\bfx_{ \hat{L},G+1} = (1,1,\ldots,1)$ based on Step 3) of the encoding procedure $\cE_{Grr}$. For $\hat{L} \leq \ell(\bfx) < \hat{L}+G$, the claim that the last $\ell(\bfx) - \hat{L} +1$ bits of $\bfx$ equal $1$ follows from Step 2) of $\cE_{Grr}$, since the string $(B(i), B(j),1,1,\ldots,1,1)$ ends with $G+2$ $1$s.
\end{IEEEproof}

\begin{proposition}\label{cl:Grec} The string $\bfx^{(k-1)}$ generated during encoding with $\cE_{Grr}$ can be reconstructed from $\bfx^{(k)},$ for all $k \geq 1$. \end{proposition}

We present next the encoder $\cE_G$ for $\cC_G(n)$. The encoder $\cE_G$ operates similarly to $\cE_{LR}$ 
in so far that it starts with an information string $\bfx_I$ of length $n$ and calls the repeat replacement encoder $\cE_{Grr}$ to remove repeated substrings. The main difference between $\cE_G$ and $\cE_{LR}$ is that $\cE_{G}$ introduces additional redundancy needed for reconstruction in the presence of coverage gaps.

\begin{algorithm}[H]
\caption{Encoder $\cE_{G}$ for an $(L,G)$-reconstruction code.}
\begin{algorithmic}[1]
  \STATE Let $\bfx_I \in \{0,1\}^{n-3G-3}$. Append $({({\bfx}_{I})_{\hat{L},G+1}, 1)} \in \{0,1\}^{G+2}$ and prepend $(1,1,\ldots,1,0) \in \{0,1\}^{G+1}$ to $\bfx_I$. Set ${(\bfx_{I})_{\hat{L}, G+1}} = (1,1,\ldots,1)$.
  \STATE If $\ell(\cE_{Grr}(\bfx_I)) = n$, set $\bfx = \cE_{Grr}(\bfx_I)$. Otherwise, append to $\cE_{Grr}(\bfx_I)$ as many $0$s as needed to make the string have fixed length $n$.
\end{algorithmic}
\end{algorithm}

The next proposition is a consequence of the description of $\cE_{Grr}$ and $\cE_{G}$.

\begin{proposition}\label{cl:EG} Suppose that $\bfx = ({\bfx}', {\bf0}) = \cE_G({\bfx}_I),$ where $\bfx'$ is the output of the encoder $\cE_{Grr}$. Then, $\bfx'$ is $\hat{L}$-substring unique, $\bfx_{1, G+1} = (1,1,\ldots,1,0)$, and the last bit of $\bfx'$ equals $1$. If $\ell(\bfx') \geq \hat{L}+G$, then $\bfx_{\hat{L},G+1} = (1,1,\ldots,1)$. Otherwise, if $\ell(\bfx') < \hat{L}+G$, then $\bfx_{\hat{L},\ell(\bfx')- \hat{L} +1} = (1,1,\ldots,1)$.   \end{proposition}

\begin{IEEEproof} The fact that $\bfx'$ is $\hat{L}$-substring unique follows using the same proof techniques described for the repeat replacement encoder $\cE_{RR}$ in Section~\ref{sec:g2logn}. The fact that $\bfx_{1, G+1} = (1,1,\ldots,1,0)$ follows from $\hat{L} > G+1$ and the observation that given any string $\bfv \in \{0,1\}^{\hat{L}-1}$, the string $(\bfx_{1, G+1}, \bfv)$ is $\hat{L}$-substring unique, and hence no repeat replacement is performed on $\bfx_{1,G+1}$. The last statement of the proposition follows directly from Step 2) of $\cE_G$ and Proposition~\ref{cl:Grr}.\end{IEEEproof}
%

\begin{lemma} The code $\cC_G(n) = \{ \bfx : \bfx = \cE_G(\bfx_I) \}$ is an $(L,G)$-reconstruction code. \end{lemma}
\begin{IEEEproof} We need to show that one can recover $\bfx$ given a  multispectrum $\widehat{M}_L(\bfx)$ with maximal coverage gap $G$. 

Let $\bfx = (\bfx', \bf0),$ where $\bfx' $ is the output of the gap repeat replacement encoder in Step 2) of $\cE_{Grr}$. We start as in Proposition~\ref{cl:uniquesubs2}, by identifying a proper prefix of $\bfx$, and then continue by appending substrings to the prefix until the codeword $\bfx$ is recovered.

Suppose that $\bfx^{(1)}$ is such that the first $\hat{L}$ bits of $\bfx^{(1)}$ appear in $\widehat{M}_L(\bfx)$ exactly once. We show that such a string $\bfx^{(1)}$ necessarily exists. Suppose that $\bfx_{j,L},$ where $1 \leq j \leq G+1,$ is the positionally first substring of length $L$ in $\bfx$ that belongs to $\widehat{M}_L(\bfx)$. Assume next that $\ell(\bfx') \geq \hat{L} + G$. We show that the substring $\bfx_{j,\hat{L}}$ appears exactly once in $\widehat{M}_{\hat{L}}(\bfx)$, which implies that $\bfx_{j,\hat{L}}$ appears exactly once in $\widehat{M}_L(\bfx)$. {Based on Proposition~\ref{cl:EG},} and given that $\bfx'$ is $\hat{L}$-substring unique, $\bfx_{j,\hat{L}}$ cannot appear twice in $\widehat{M}_{\hat{L}}(\bfx')$. Therefore, if $\bfx_{j,\hat{L}}$ appears twice in $\widehat{M}_{\hat{L}}(\bfx)$, one must have $\bfx_{j,\hat{L}} \in \{ \bfx_{\ell(\bfx')-\hat{L}+2}, \bfx_{\ell(\bfx')-\hat{L}+3},$ $\ldots,$ $\bfx_{n-\hat{L}+1, \hat{L}} \}$. However, this is not possible as each string in $\{ \bfx_{\ell(\bfx')-\hat{L}+2}, \bfx_{\ell(\bfx')-\hat{L}+3},$ $\ldots,$ $\bfx_{n-\hat{L}+1,\hat{L}} \}$ ends with a $0$, while given $\ell(\bfx') \geq \hat{L} + G$, the string $\bfx_{j,\hat{L}}$ must end with a $1$ according to Proposition~\ref{cl:EG}.

Next, we consider the case $\ell(\bfx') < \hat{L} + G$. If $\bfx_{j, \hat{L}}$ ends with a $1$, then the result follows based on the same argument as used for the case $\ell(\bfx') \geq \hat{L} + G$. Assume instead that $\bfx_{j, \hat{L}}$ ends with a $0$. This is possible only if $(j-1) + \hat{L} > \ell(\bfx'),$ since $\bfx'$ ends with a $1$. Let $k$ be the position of the last $1$ in the substring $\bfx_{j, \hat{L}}$. Then, the strings $\{ \bfx_{j+1, \hat{L}}, \bfx_{j+2, \hat{L}}, \ldots, \bfx_{j+(k-1), \hat{L}} \}$ all have their last $1$ appear at positions $< k$. Since the strings in $\widehat{M}_{\hat{L}}(\bfx)/\{ \bfx_{j+1, \hat{L}}, \bfx_{j+2, \hat{L}}, \ldots, \bfx_{j+(k-1), \hat{L}} \}$ are all-zeros, there exists a string $\widehat{M}_L(\bfx)$ whose first $\hat{L}$ bits appear only once in $\widehat{M}_{\hat{L}}(\bfx)$. This settles the case $\ell(\bfx') < \hat{L} + G$.

Recall that $\bfx^{(1)}$ is such that the substring $\bfx^{(1)}_{1,\hat{L}}$ appears exactly once in $\widehat{M}_L(\bfx)$; we established the existence of such a string with the previous argument. It is straightforward to show that $\bfx^{(1)} = \bfx_{j, L},$ where $1 \leq j \leq G+1$ and $\bfx_{j,L}$ is the positionally first substring of $\bfx$ of length $L$ that belongs to $\widehat{M}_L(\bfx)$. Our goal is to find a prefix of $\bfx$ given $\bfx^{(1)}$. Suppose next that the first $0$ in $\bfx^{(1)}$ appears at position $k$. Since $\bfx_{1,G+1} = (1,1,\ldots,1,0),$ based on Proposition~\ref{cl:EG}, one can form $\bfx_{1,L+G-k+1}$ by prepending $G-k+1$ $1$s to $\bfx^{(1)}$. Let $\hat{\bfx} = \bfx_{1,L+G-k+1}$, so that $\hat{\bfx}$ is a prefix of $\bfx$ of length $L+G-k+1$. We continue with a procedure similar to that outlined in the proof of Proposition~\ref{cl:uniquesubs2} and extend the string $\hat{\bfx}$. For $j > (L+G-k+1)-L+1$, we attempt to find a substring $\bfx_L$ such that ${(\bfx_{L}})_{1,\hat{L}} = \hat{\bfx}_{j, \hat{L}}$, and $j$ is as small as possible. We note that since $\widehat{M}_L(\bfx)$ has maximal coverage gap $G$, such an $\bfx_L$ exists as there are $(L-1) - \hat{L} + 1 = G+1$ possibilities for $j$. Also, observe that there exist at most two substrings $\bfx_L$ and $\bfx_L'$ that satisfy ${(\bfx_{L})}_{1,\hat{L}} = \hat{\bfx}_{j, \hat{L}}$, ${(\bfx'_{L})}_{1,\hat{L}} = \hat{\bfx}_{j, \hat{L}},$ unless $\bfx_L, \bfx'_L$ are both all-zeros. The latter observation holds since $\bfx'$ is $\hat{L}$-substring unique, so that $\hat{\bfx}_{j, \hat{L}}$ cannot appear more than once in $\widehat{M}_{\hat{L}}(\bfx')$. Assume therefore that $\hat{\bfx}_{j,\hat{L}}$ appears in $\{ \bfx_{\ell(\bfx')-{L}+2, \hat{L}}, \bfx_{\ell(\bfx')-{L}+3, \hat{L}}, \ldots, \bfx_{n-{L}+1, \hat{L}} \}$. If $\hat{\bfx}_{j, \hat{L}}$ is not all-zeros it follows that $\hat{\bfx}_{j, \hat{L}} \in \{ \bfx_{\ell(\bfx')-\hat{L}+2, \hat{L}}, \bfx_{\ell(\bfx')-\hat{L}+3, \hat{L}},$ $\ldots,$ $\bfx_{\ell(\bfx'), \hat{L}} \}$. However, similarly to what was previously discussed, since $\bfx'$ ends with a $1$ and is followed by an all-zeros string in $\bfx$, $\hat{\bfx}_{j, \hat{L}}$ cannot appear more than once in $\{ \bfx_{\ell(\bfx')-\hat{L}+2, \hat{L}}, \bfx_{\ell(\bfx')-\hat{L}+3, \hat{L}},$ $\ldots,$ $\bfx_{\ell(\bfx'), \hat{L}, \hat{L}} \}$ -- otherwise, the last $1$ in each string would appear at a different position.

As a result, there exist at most two substrings, say $\bfx_L$ and $\bfx_L'$ such that ${(\bfx_{L})}_{1,\hat{L}} = \hat{\bfx}_{j, \hat{L}}$, ${(\bfx_L')}_{1,\hat{L}} = \hat{\bfx}_{j, \hat{L}},$ unless $\bfx_L$ and $\bfx'_L$ are both all-zeros. If $\bfx_L$ and $\bfx_L'$ are all-zeros, we append $j-\left( (L+G-k+1)-L+1 \right)$ $0$s to $\hat{\bfx}$. Suppose then that $\bfx_L$ and $\bfx_L'$ are not both equal to the all-zeros string. Then, at least one of the strings $\bfx_L$ or $\bfx_L'$ ends with $j-G+k-2$ $0$s, since it has to belong to the set $\{ \bfx_{\ell(\bfx')-{L}+2, {L}}, \bfx_{\ell(\bfx')-{L}+3, {L}},$ $\ldots,$ $\bfx_{\ell(\bfx'), {L}, {L}} \}$. Assume that the string $\bfx_L$ ends with $j-G+k-2$ $0$s. Then, given that $\bfx_L \neq \bfx'_L$, it follows that the last $j-G+k-2$ bits of $\bfx_L'$ are not all equal to $0$. In this case, we extend the string $\hat{\bfx}$ by appending the last $j-G+k-2$ bits of $\bfx_L'$ to $\hat{\bfx}$ and remove the string $\bfx_L'$ from $\widehat{M}_L(\bfx)$, so that $\hat{\bfx}$ is a prefix of $\bfx$ of length $L+j-1$. We continue in this manner until either the set $\widehat{M}_L(\bfx)$ is empty or $\widehat{M}_L(\bfx)$ contains only all-zeros strings. Then, we recover the string $\bfx$ from $\hat{\bfx}$ by appending an appropriate number of $0$s.
\end{IEEEproof}

The next lemma follows from Proposition~\ref{cl:Grec}.

\begin{lemma} Given $\bfx = \cE_G(\bfx_I)$, one can recover $\bfx_I$. \end{lemma}

As a consequence of the two previous lemmas, we have the following result.

\begin{theorem} For a positive integer $G$, there exists an $(2 \log n + 2G + 4,G)$ reconstruction code of size $2^{n-3G-3}$. \end{theorem}

\subsection{Codes for Spurious and Coverage Errors}

We now turn our attention to the problem of correcting both spurious and coverage errors. In order to describe the problem, we require some additional notation. 

Let $\cE_t : \{0,1\}^{m} \to \{0,1\}^m$ be a map such that the input and output vector differ in at most $t$ positions. In other words, $\cE_t$ is such that given any $\bfv \in \{0,1\}^m$, $d_H(\bfv, \cE_t(\bfv)) \leq t,$ where $d_H(\bfv, \cE_t(\bfv))$ denotes the Hamming distance between $\bfv$ and $\cE_t(\bfv)$. As before, for $\bfx \in \{0,1\}^n,$ let $M_L(\bfx)$ denote its $L$-multispectrum and $\widehat{M}_L(\bfx) = \{ \bfx_1, \bfx_2, \ldots, \bfx_M \} \subseteq M_L(\bfx)$ has maximal coverage gap $G$. We say that $\widetilde{M}_L(\bfx)$ is $(G,t)$-constrained if $\widetilde{M}_L(\bfx) = \{ \cE_t(\bfx_1), \cE_t(\bfx_2), \ldots, \cE_t(\bfx_M) \}$. 

\begin{example} Let $\bfx = (0,1,1,0,1,0,0)$. Then, $M_4(\bfx) = \Big \{ (0,1,1,0), (1,1,0,1), (1,0,1,0), (0,1,0,0) \Big \}$. 
Assume that the observed spectrum with coverage errors equals $\widehat{M}_4(\bfx) = \Big \{ (1,0,1,0), (0,1,0,0) \Big \}$. Clearly, $\widehat{M}_4(\bfx)$ has maximal coverage gap equal to two. The set $\widetilde{M}_4(\bfx) = \Big \{ (1,\textcolor{red}{1},1,{0}), (0,\textcolor{red}{0},0,0) \Big \}$ is $(2,1)$-constrained. 
\end{example}

Note that the substrings in $\widetilde{M}_4(\bfx)$ cover the bit $x_4$ twice. The first copy of $x_4$ appears as the second bit in $(1,1,1,0)$, while the second copy of $x_4$ appears as the first bit in $(0,0,0,0)$; $x_4$ was subjected to a substitution error in the string $(1,1,1,0)$ but not in the string $(0,0,0,0)$. A spectrum $\widetilde{M}_L(\bfx)$ is said to be \emph{reliable} if for any symbol in $\bfx$, there are more copies of the correct value rather than incorrect value of the symbol.

\begin{example} Let $\bfx = (0,1,1,0,1,0,0,1,0,0,0,1)$. Then, 
$$M_7(\bfx) = \Big \{ (0,1,1,0,1,0,0), (1,1,0,1,0,0,1), (1,0,1,0,0,1,0), (0,1,0,0,1,0,0), (1,0,0,1,0,0,0), (0,0,1,0,0,0,1)\Big \}.$$ 
Assume that coverage errors lead to
$$\widehat{M}_7(\bfx) = \Big \{ (0,1,1,0,1,0,0), (0,1,0,0,1,0,0), (1,0,0,1,0,0,0)\Big \}.$$ 
The noisy spectrum $\widehat{M}_7(\bfx)$ has maximal coverage gap two, since ${M}_7(\bfx) \setminus \widehat{M}_7(\bfx) = \{ \bfx_{1,7}, \bfx_{4,7}, \bfx_{5,7} \}$. Suppose next that 
$$\widetilde{M}_7(\bfx) = \Big \{ (0,1,1,0,1,0,\textcolor{red}{1}), (0,1,0,0,1,0,0), (1,0,0,1,0,0,0)\Big \}.$$ 
It is straightforward to see that $\widetilde{M}_7(\bfx)$ is $(2,1)$-constrained. Furthermore, $\widetilde{M}_7(\bfx)$ is reliable. To see why this is the case, note that only one single substitution error occurs in $\widetilde{M}_7(\bfx)$, within the substring $(0,1,1,0,1,0,0)$. The error affects the bit $x_7$, which appears at the seventh position in the substring. However, the spectrum also contains two additional correct copies of the bit $x_7$ that appear at the fourth position in $(0,1,0,0,1,0,0)$ and at the third position in $(1,0,0,1,0,0,0)$.
\end{example}

We say that a code is an $(L,G,t)$-error correcting code if its codewords can be uniquely reconstructed given a noisy spectrum $\widetilde{M}_L(\bfx)$ that is $(G,t)$-constrained and reliable. In the remainder of this section, we describe a construction for a $(L,G,t)$-error correcting code $\cC_{G,t}(n)$. 

The main difference between the error model discussed in this section and the coverage-error model described in the previous section is that errors may occur in different copies of the data bits covered by substrings in $\widetilde{M}_L(\bfx)$. To deal with this issue, we introduce a set of new coding constraints. To define the constraints, let $\widetilde{L}$ be a positive integer whose value is determined based on the parameters $G < \widetilde{L}$ and $t$. In addition, suppose that $L = 3 \widetilde{L}$. As before, a codeword $\bfx \in \cC_{G,t}(n)$ is assumed to take the form $\bfx = (\bfx', {\bf0})$. The constraints of interest are imposed on the string $\bfx'$ as follows: 
\begin{itemize}
\item \textbf{Condition 1:} Any two distinct substrings of length $\widetilde{L}$ in $\bfx'$ are required to be at Hamming distance $\geq 6t+1$; 
\item \textbf{Condition 2:} Any substring of $\bfx'$ of length $\widetilde{L}$ has weight $\geq 2t+1$. 
\end{itemize}
If $\bfx'$ satisfies Condition 1, we say that $\bfx'$ is $(6t+1)$-substring distinct, and if $\bfx'$ satisfies Condition 2, we say that $\bfx'$ is $(2t+1)$-heavy. We will also require three additional constraints to be described shortly after introducing some relevant notation. 

Let $Enc_N : \{0,1\}^{N- (5t+G)\log N} \to \{0,1\}^{(5t +G) \log N}$ be a systematic encoder for a code of length $N$ and minimum distance $2(5t+G)+1$ that has $(5t+G) \log N$ redundant bits. In other words, for any two distinct binary strings $\bfz_1, \bfz_2$ of the same length, $d_H \bigg ( \Big (\bfz_1, Enc_{\ell(\bfz_1)}(\bfz_1) \Big), \Big (\bfz_2, Enc_{\ell(\bfz_2)}(\bfz_2) \Big ) \bigg)  \geq 2(5t+G) + 1$. The encoder $Enc_N$ will be used to protect the last ${L}-1$ bits of $\bfx'$ along with the information provided by $B\left(\ell(\bfx') \bmod (2(G+\widetilde{L}+1) )\right)$, {the previously used binary representation function}. For convenience, reference to the parameter $N$ is omitted whenever it is clear from the context what the length of the input string fed into $Enc_N$ is. 

For convenience, let 
$$\bfx_{t,\text{rep}} = (1^{2t+G+1}, 0, 1^{2t+G+1}, 0, \ldots, 1^{2t+G+1}, 0) \in \{0,1\}^{(2t+G+2)(2t+2)}$$ 
denote a string obtained by concatenating $(1_{2t+G+1}, 0) \in \{0,1\}^{2t+G+2}$ with itself $2t+2$ times. Furthermore, let $P_L$ be a positive integer that satisfies $P_L - (5t+G) \log P_L -(2t+G+2)(2t+2) = \log (2(G+\widetilde{L}+1))$. Finally, let 
$$\bfv_{\ell(\bfx')} =\Bigg ( B\Big(\ell(\bfx') \bmod (2(G+\widetilde{L}+1)) \Big), Enc\Big(B\Big(\ell(\bfx') \bmod (2(G+\widetilde{L}+1)) \Big) \Bigg ) \in \{0,1\}^{P_L-(2t+G+2)(2t+2)}.$$
Note that $\ell(\bfx') \bmod (2(G+\widetilde{L}+1))$ can be recovered from the string $\cE_{5t+G}(\bfv_{\ell(\bfx')})$.

The additional constraints required are used to recover the bits forming a sufficiently long prefix and suffix of $\bfx'$, and are summarized as follows:
\begin{itemize}
\item \textbf{Constraint 4:} The last symbol of $\bfx'$ is equal to $1$;
\item \textbf{Constraint 5:} The last ${L}-1$ bits of $\bfx'$ constitute a codeword of a code with minimum Hamming distance $2(5t+G) + 1$;
\item \textbf{Constraint 6:} The substring $\bfx_{1, P_L}$ is equal to $(\bfx_{t,\text{rep}}, \bfv_{\ell(\bfx')})$.
\end{itemize}
Constraint 5 will be used in Proposition~\ref{cl:ends2} to recover a prefix of $\bfx$ used to jumpstart the reconstruction procedure. All three constraints are used to determine a (possibly noisy) suffix of $\bfx'$.

As before, we make use of two encoders. The first encoder, $\cE_{GTrr}$, is what we term a $(G,t)$-repeat replacement encoder, and it represents a variant of the basic repeat replacement encoder. The encoder $\cE_{GTrr}$ outputs a string free of repeats of predetermined length. However, unlike all other repeat replacement encoders introduced so far, $\cE_{GTrr}$ also ensures that its output is $(6t+1)$-substring distinct, $(2{t}+1)$-heavy, and that the last ${L}-1$ bits form a codeword from a code with minimum Hamming distance $2(5t+G)+1$ that ends with a $1$, and that $\bfx_{1, P_L} = (\bfx_{t,\text{rep}}, \bfv_{\ell(\bfx')} )$. The second encoder, $\cE_{GT}$, operates on the output of the encoder $\cE_{GTrr}$ to form the codewords of the desired code.

We now rigorously describe the encoding steps of $\cE_{GTrr}$. Let $\widetilde{L}$ be a positive integer such that $\widetilde{L} = 2 \log n + 6t \log( \widetilde{L}+1) + (5t+G) \log ({L}-1) + P_L + 2t + 2$, and suppose that $L = 3 \widetilde{L}$. In what follows, we assume that $G < \widetilde{L}$. We say $\bfz \in \{0,1\}^m$ has an $(\widetilde{L}, 6t)$-repeat at positions $(i,j)$ if $d_H(\bfz_{i, \widetilde{L}}, \bfz_{j, \widetilde{L}} ) \leq 6t$. 

Furthermore, let $D(\bfz_1, \bfz_2)$ be a mapping that takes as its input two strings $\bfz_1, \bfz_2,$ 
{both of length $\widetilde{L}$,} that differ in at most $6t$ positions, and outputs a string of length $6t \log (\widetilde{L}+1)$ describing the positions where the two strings differ. Clearly, given $\bfz_1$ or $\bfz_2$ and $D(\bfz_1, \bfz_2)$, one can uniquely determine $\bfz_2$ or $\bfz_1$.

The input to the desired repeat replacement encoder $\cE_{GTrr}$ is a string $\bfx_I \in \{0,1\}^n$ whose prefix of length $P_L$ equals $(\bfx_{t,\text{rep}}, \bfv_{n} )$, its last bit has the value $1$, and ${(\bfx_{I})}_{n- {L}+2, {L}-1}$ belongs to a code with minimum Hamming distance $2(5t+G)+1$. 

\begin{algorithm}[H]
\caption{Repeat replacement encoder $\cE_{GTrr}$ for generating $(6t+1)$-substring distinct strings.}
\begin{algorithmic}[1]
\STATE If $\bfx_I$ is $(6t+1)$-substring distinct and $(2t+1)$-heavy, set $\bfx = \bfx_I$ and STOP. Otherwise, set $\bfx^{(0)} = \bfx_I$, and let $k=1$.
\STATE If $\bfx^{(k-1)}$ has a $(\widetilde{L}, 6t)$-repeat at positions $(i,j)$, go to Step 3. Otherwise, go to Step 4.
\STATE Let 
$$\bfv = \Bigg (B(i), B(j), D( {\bfx}^{(k-1)}_{i,\widetilde{L}}, {\bfx}^{(k-1)}_{j,\widetilde{L}}), 1_{P_L}, 1_{2t+1} \Bigg) \in \{0,1\}^{\widetilde{L}-1-(5t+G)\log({L}-1)}.
$$ 
{Go to Step 5.}
\STATE Let $2t$ be an upper bound on the weight of $\bfx^{(k-1)}_{j, \widetilde{L}}$ and let
$$\bfv = \Bigg (B(j), B(j), D( {\bf0}, {\bfx}^{(k-1)}_{j,\widetilde{L}}), 1_{P_L}, 1_{2t+1} \Bigg) \in \{0,1\}^{\widetilde{L}-1-(5t+G)\log({L}-1)}.$$ 
{\STATE Let $\bfx^{(k)}$ be the string obtained by deleting the substring $\bfx^{(k-1)}_{j, \widetilde{L}}$ from $\bfx^{(k-1)}$.} 

\STATE Update the last $2t + 1 + P_L$ bits of $\bfv$ to $(\bfx^{(k)}_{1, P_L}, 1_{2t+1})$ and append $(Enc(\bfx^{(k)}_{\ell(\bfx^{(k)})-2\widetilde{L}+1, 2\widetilde{L}}, \bfv), \bfv) \in \{0,1\}^{\widetilde{L}-1}$ to $\bfx^{(k)}$. Update 
$$\bfx^{(k)}_{1, P_L}= (\bfx_{t,\text{rep}}, \bfv_{\ell(\bfx^{(k)})} ).$$
If $\bfx^{(k)}$ is $(6t+1)$-substring distinct and $(2t+1)$-heavy, set $\bfx= \bfx^{(k)}$ and STOP. Otherwise, set $k=k+1$, and go to Step 2.
\end{algorithmic}
\end{algorithm}
In Step 5 of $\cE_{GTrr}$, if $\ell(\bfx^{(k)}) < 2\widetilde{L}$, then $\bfx^{(k)}_{\ell(\bfx^{(k)})-2\widetilde{L}+1, 2\widetilde{L}}$ (provided as input to $Enc$) is the string obtained by appending sufficiently many $0$s to $\bfx^{(k)}$ so that the resulting string length equals $2\widetilde{L}$.

Also, observe that if Step 3 is executed, $\bfv$ starts with $B(i), B(j),$ where $i \neq j$; otherwise, if Step 4 is executed, $\bfv$ starts with $B(j), B(j)$. The next two propositions easily follow using previously described proof techniques.

\begin{proposition}\label{cl:Grr} The output string $\bfx$ has the following properties:
\begin{enumerate}
\item $\bfx$ is $(6t+1)$-substring distinct;
\item $\bfx$ is $(2t+1)$ heavy;
\item $\bfx$ ends with a $1$;
\item The suffix of $\bfx$ of length ${L}-1$ is a codeword in a code with minimum Hamming distance $2(5t+G)+1$; and
\item $\bfx_{1, P_L} =  (\bfx_{t,\text{rep}}, \bfv_{\ell(\bfx)} )$.
\end{enumerate} \end{proposition}

\begin{proposition}\label{cl:gtrecover} For $k\geq 1$, any string $\bfx^{(k-1)}$ generated by $\cE_{GTrr}$ can be reconstructed from $\bfx^{(k)}$.  \end{proposition}

The encoder $\cE_{GT}$ described next uses the repeat replacement encoder $\cE_{GTrr}$ to generate the codewords in $\cC_{G,T}(n)$.

\begin{algorithm}[H]
\caption{The encoder $\cE_{GT}$ of a $(L,G,t)$-reconstruction code.}
\begin{algorithmic}[1]
  \STATE Let $\bfx_I \in \{0,1\}^{n-P_L-(5t+G)\log({L}-1) - 1}$ and suppose that $\bfv$ is the suffix of $\bfx_I$ of length \\
  ${L}-2- (5t+G)\log({L}-1)$. Append $(Enc(\bfv,1),\bfv,1)$ and prepend $(\bfx_{t,\text{rep}}, \bfv_{n} )$ to $\bfx_I$.
  \STATE Append to $\cE_{GTrr}(\bfx_I)$ as many $0$s as needed to make the string have length $n$.
\end{algorithmic}
\end{algorithm}

Next, we turn our attention to describing the reconstruction algorithm, $\widetilde{\cR} : \widetilde{M}_L(\bfx) \to \bfx$, for the $(L,G,t)$-reconstruction code $\cC_{G,t}(n)$. We start by introducing some relevant notation. 

For two strings $\bfz \in \{0,1\}^{m}$, $m \geq L$, and $\bfx_L \in \widetilde{M}_L(\bfx)$, we say that $\bfx_L$ is a $6t$-distant suffix of $\bfz$ at position $j$ with \emph{overhang} $h$ if $d_H \big({(\bfx_{L})}_{1,\widetilde{L}}, \bfz_{j, \widetilde{L}} \big ) \leq 6t,$ for some $m-L+1 \leq j \leq m-\widetilde{L}+1$ {and $h=L-(m-j+1)$}. {Intuitively, if $\bfx_L$ is a $6t$-distant suffix of $\bfz$ at position $j$ with \emph{overhang} $h$, then the first $\widetilde{L}$ bits of $\bfx_{L}$ are within Hamming distance $6t$ of $\bfz_{j, \widetilde{L}}$. In addition, if those first $\widetilde{L}$ bits of $\bfx_L$ are aligned with the suffix of $\bfz$ beginning at position $j$, then the final $h$ bits of $\bfx_L$ extend beyond the end of $\bfz$ (or overhang $\bfz$) by $h$ bits.} If $\bfx_L$ is a $6t$-distant suffix with overhang $h$ at a position $j$ that is the largest such possible, then we say that $\bfx_L$ is a furthest $6t$-distant suffix of $\bfz$ in $\widetilde{M}_L(\bfx)$. We will also refer to $\bfz$ as a $6t$-prefix of $\bfx_L$. Furthermore, for any $\bfx_L \in \widetilde{M}_L(\bfx)$, we say that $\bfx_L$ is $(t+1)$-tail heavy if ${(\bfx_{L})}_{2\widetilde{L}+1, \widetilde{L}}$ is $(t+1)$-heavy. 

Suppose that $\widetilde{\bfx}$ is a noisy copy of $\bfx$ containing substitution errors. For $\bfx_L \in \widetilde{M}_L(\bfx)$, we say that ${(x_{L})_j}$ is an \emph{approximate repeat} of $\widetilde{x}_{i}$ if $d_H({(\bfx_{L})}_{1, \widetilde{L}}, \widetilde{\bfx}_{i-j+1, \widetilde{L}}) \leq 6t$ and ${\bfx}_L$ is $(t+1)$-tail heavy.

\begin{algorithm}[H]
\caption{Decoding algorithm $\widetilde{\cR}$ for the code $\cC_{G,t}(n)$.}
\begin{algorithmic}[1]
  \STATE \textbf{\textit{Initialization:}} If there exists a $(t+1)$-tail heavy $\bfx_L \in \widetilde{M}_L(\bfx)$ that has no $6t$-distant prefixes in $\widetilde{M}_L(\bfx)$, set $\widetilde{\bfx}^{(0)} = \bfx_{L}$. Otherwise, let $\widetilde{\bfx}^{(0)}$ be a string $\bfx_L \in \widetilde{M}_L(\bfx)$ such that ${(\bfx_{L})}_{j, 2t+1}$ is $(t+1)$-heavy and $j$ is maximal.
  
  \STATE For $i \in [2t+G+2]$, set $\widetilde{x}^{(0)}_i$ to the majority-value of the multiset $\{ \widetilde{x}^{(0)}_{i}, \widetilde{x}^{(0)}_{i+(2t+G+2)}, \widetilde{x}^{(0)}_{i+2(2t+G+2)}, \ldots, \widetilde{x}^{(0)}_{i + (2t)(2t+G+2)}\}$. Suppose that the first $0$ in $\widetilde{\bfx}^{(0)}$ appears at position $2t + G + 2 - j$. Then, let $\widetilde{\bfx}^{(1)}$ be the string obtained by prepending $j$ $1$s to $\widetilde{\bfx}^{(0)}$. Furthermore, let $\ell = \ell(\bfx') \bmod (2(G + \widetilde{L}+1))$, which can be recovered from the first $P_L$ symbols of $\widetilde{\bfx}^{(1)}$. 
  
  \STATE \textbf{\textit{Extension:}} Let $\bfx_L \in \widetilde{M}_L(\bfx)$ be a furthest $(t+1)$-tail heavy $6t$-distant suffix of $\widetilde{\bfx}^{(k)}$ with overhang $h$. Let $\widetilde{\bfx}^{(k+1)}$ be the result of appending the last $h$ bits of $\bfx_L$ to $\widetilde{\bfx}^{(k)}$. If no such $\bfx_L$ exists, go to Step 4. Otherwise, update $k=k+1$, and repeat this step.
  
  \STATE \textbf{\textit{Length adjustment:}} If $0 < (\ell(\widetilde{\bfx}^{(k)})-\ell) \bmod 2(G+\widetilde{L}+1) \leq \widetilde{L}-1$, let $\widetilde{\bfx}$ be a truncation of $\widetilde{\bfx}^{(k)}$ such that $(\ell(\widetilde{\bfx}^{(k)})-\ell) \bmod 2(G+\widetilde{L}+1) = 0$. If $\widetilde{L}-1 < (\ell(\widetilde{\bfx}^{(k)})-\ell) \bmod 2(G+\widetilde{L}+1) \leq 2(G+\widetilde{L} + 1) -1$, let $\widetilde{\bfx}$ be the string obtained by appending $0$s to $\widetilde{\bfx}^{(k)}$ until $(\ell(\widetilde{\bfx}^{(k)})-\ell) \bmod 2(G+\widetilde{L}+1) = 0$.  
  
  \STATE \textbf{\textit{Error correction:}} For $i \in [\ell(\widetilde{\bfx})-{L}+1]$, set $\widetilde{x}_{i}$ to the majority value of the approximate repeats of $\widetilde{x}_{i}$ covered by substrings in $\widetilde{M}_L(\bfx)$.
   
  \STATE \textbf{\textit{Termination:}} Using the decoder for the length ${L}-1$ codebook, recover $\bfx'_{\ell(\bfx')-{L}+2, {L}-1}$ and set $\widetilde{\bfx}_{\ell(\bfx')-L+2, L-1} = \bfx'_{\ell(\bfx')-L+2, L-1}$, $\overline{\bfx} = (\widetilde{\bfx}, \bf0)$.
\end{algorithmic}
\end{algorithm}

We break down the proof of correctness of the decoding procedure $\widetilde{\cR}$ into several steps.

\begin{proposition} After Step 1, if $\ell(\bfx') \geq L + G$, then $\widetilde{\bfx}^{(0)} = \cE_t(\bfx_{j,L}),$ where $1 \leq j \leq G+1$. Otherwise, if $\ell(\bfx') < L + G$, then $\widetilde{\bfx}^{(0)} = \cE_t(\bfx_{j,L}),$ where $1 \leq j \leq 2t + G + 1$.  \end{proposition}
\begin{IEEEproof} Since $\widetilde{M}_L(\bfx)$ is $(G,t)$-constrained, we know that at least one of the strings $\cE_t(\bfx_{1,L}), \cE_t(\bfx_{2,L}), \ldots, \cE_t(\bfx_{G+1,L})$ belongs to the set $\widetilde{M}_L(\bfx)$. It is straightforward to verify that if $\ell(\bfx') \geq L+G$, then $\widetilde{\bfx}^{(0)} = \cE_t(\bfx_{j,L}),$ where $1 \leq j \leq G+1,$ since each of the strings in $\cE_t(\bfx_{1,L}), \cE_t(\bfx_{2,L}), \ldots, \cE_t(\bfx_{G+1,L})$ is $(t+1)$-tail heavy. 

If $\ell(\bfx') < L + G$, then the result follows by observing that in order for ${(\bfx_{L})}_{j, 2t+1}$ to be $(t+1)$-heavy, ${(\bfx_{L})}_{j,2t+1} \neq \cE_t(0,0,\ldots,0)$, since $\cE_t$ can introduce at most $t$ $1$s into its argument strings. In addition, given that the output string of the encoder $\cE_{GTrr}$, $\bfx'$, ends with $2t+1$ $1$s, there exists a $\bfx_L \in \widetilde{M}_L(\bfx)$ for which ${(\bfx_{L})}_{j, 2t+1}$ is $(t+1)$-heavy and ${(\bfx_{L})}_{j, 2t+1}$ contains a copy of at least one symbol from $\bfx'$. In the worst case, we have ${(\bfx_{L})}_{j, 2t+1} = \cE_t(x'_{\ell(\bfx')}, 0,0,\ldots,0)$. Since $\widetilde{\bfx}^{(0)}=\bfx_L$ was chosen so that $j$ is maximal and $\widetilde{M}_L(\bfx)$ is $(G,t)$-constrained, it follows that $\bfx^{(0)} = \cE_t(\bfx_{k,L})$ for all $1 \leq k \leq 2t + G + 1,$ as claimed. \end{IEEEproof}

\begin{proposition}\label{cl:ends2} Let $m = \min\{ \ell(\widetilde{\bfx}^{(1)}), \ell(\bfx') \}$. After Step 2, $$d_H( \widetilde{\bfx}^{(1)}_{1, m}, \bfx'_{1,m}) \leq t.$$ \end{proposition}
\begin{IEEEproof}  As a result of the previous proposition, we have $\widetilde{\bfx}^{(0)} = \cE_t(\bfx_{j,L})$ for all $1 \leq j \leq 2t + G + 1$. Thus, $\widetilde{\bfx}^{(0)}_{1, (2t+G+2)(2t+1)} = \cE_t( 1_{2t+G+2-j}, 0,1_{2t+G+1}, 0, \ldots, 1_{2t+G+1}, 0, 1_{j-1})$. In words, $\widetilde{\bfx}^{(0)}_{1, (2t+G+2)(2t+1)}$ is the result of at most $t$ errors occurring to a cyclic rotation of the string $(1_{2t+G+1}, 0)$ concatenated $(2t+1)$ times. Since $\cE_t$ introduces at most $t$ errors, we can correct any errors that occur in the first $(2t+G+2)$-symbols of $\widetilde{\bfx}^{(0)}$. Subsequently, we can prepend symbols to $\widetilde{\bfx}^{(0)}$ to obtain $\widetilde{\bfx}^{(1)}$, which is a string that has the properties stated in the proposition.
\end{IEEEproof}

\begin{proposition} For iteration $k$ executed during Step 3 of the algorithm, let $m_k = \min \{ \ell(\widetilde{\bfx}^{(k)}), \ell(\bfx') \}$. Then, for any $i \in [m_k-L+1]$,
$$d_H(\widetilde{\bfx}^{(k)}_{i,L}, \bfx'_{i,L}) \leq 5t.$$
\end{proposition}

\begin{IEEEproof} Suppose that at the beginning of Step 4, $k=T$. We first show the result holds for $1 \leq k \leq T-1$ using induction on $k$. For $k = 1$, the result follows from Proposition~\ref{cl:ends2}. Next, suppose the result holds for all $k \leq K$ and consider the case $k=K+1$. Suppose that the vector $\bfx^{(K+1)}$ is the result of appending the last $h=\ell(\bfx^{(K+1)})- \ell(\bfx^{(K)})$ bits of $\bfx_L$ to $\bfx^{(K)}$ as described in Step 3. By the inductive assumption, for any $i \in [m_{K}-L+1]$, one has $d_H(\widetilde{\bfx}^{(K)}_{i,L}, \bfx'_{i,L}) \leq 5t.$ Since $\bfx_L$ is $(t+1)$-tail heavy, it follows that $\bfx_L \in \widetilde{M}_L(\bfx_{1, \ell(\bfx')+\widetilde{L}})$, since in order for $\bfx_L$ to be $(t+1)$-tail heavy, ${(\bfx_{L})}_{2\widetilde{L}+1,\widetilde{L}}$ must contain at least one copy of a bit from $\bfx'$. Furthermore, since $\bfx_L$ is a $6t$-distant suffix of $\widetilde{\bfx}^{(K)}$, it follows that ${(\bfx_{L})}_{L-h+1,h} = \cE_t(\bfx_{m_K+1,h})$. Finally, since $\widetilde{M}_L(\bfx)$ is $(G,t)$-constrained and $G < \widetilde{L}$, $h \geq \widetilde{L}$, the result holds for all $1 \leq k \leq T-1$. The result for $k=T$ may be established along the same lines.
\end{IEEEproof}

The proof of the next proposition follows immediately from the previous proposition and the fact that the value $\ell(\bfx') \bmod (2(G+\widetilde{L}+1))$, which can be obtained from the first $P_L$ symbols of the received string, is encoded using a code with minimum distance $2(5t+G)+1$.

\begin{proposition} After Step 4, $\ell(\widetilde{\bfx}) = \ell(\bfx')$, and for any $i \in [\ell({\bfx}')-{L}+1]$,
$$d_H(\widetilde{\bfx}_{i,L}, \bfx'_{i,L}) \leq 5t.$$
\end{proposition}

As a consequence of the previous proposition we also have the following result.

\begin{proposition} After Step 5, $\widetilde{\bfx}_i = \bfx'_i$ for $i \in [\ell({\bfx}')-{L}+1]$. \end{proposition}

Finally, we arrive at the following lemma, which follows since the last $L-1$ symbols of $\bfx'$ belong to a code with minimum Hamming distance $2(5t+G) + 1$.

\begin{lemma} After Step 6, $\overline{\bfx} = \bfx$.  \end{lemma}

We summarize the previous results as stated in the next theorem.

\begin{theorem} There exists an $(L,G,t)$-reconstruction code with at most 
$$ \log \log n + 2(5t+G) \log (L-1) + 1$$
redundant bits whenever $L = O(\log n)$.
\end{theorem}

\section{Open problems}

{The problem of designing large codebooks of substring-unique strings may be extended in multiple directions, including:
\begin{itemize}
\item Designing substring-unique codewords over larger alphabet, and in particular, alphabets of size four that would 
accommodate applications in DNA-based data storage.
\item Incorporating address and forbidden substring constraints. Addresses are added to long DNA blocks for the purpose of random access~\cite{yazdi2015,yazdi2016}. For proper information retrieval, one requires that sequences at small Hamming or edit distance be avoided in the information string. Furthermore, due to synthesis constraints, long runs of $G$ symbols in the DNA code are to be avoided. This poses the interesting question of combining substring uniqueness constraints with runlength and other relevant string content restrictions.  
\item Extending the work to account for other types of errors that are encountered in both DNA synthesis and sequencing. In the former case, one would encounter bursts of substring errors, in which overlapping substrings would share the 
same error. A particularly interesting question is to address the reconstruction problem without the strong "coverage gap" constraint, which requires that not more than a small constant number of consecutive substrings are unobserved.
\item Accommodating deletion errors in the read and write process. This is a particularly challenging question, as one would require that no two substrings of a string be at small Levenshtein distance from each other.
\end{itemize}
\vspace{0.1in}

\hspace{-0.2in} \textbf{Acknowledgment:} The authors gratefully acknowledge funding from the NSF grant CCF 16-18366 and the Center for Science of Information (CSoI), an NSF Science and Technology Center, under grant agreement CCF-0939370.

\begin{appendices}
\section{Proof of Lemma~\ref{lem:Unb}}

To derive the lower bound, we establish an upper bound on the number of strings that do not belong to $\cU(n,L)$. We start with a string $\bfx$ of length $n-L+1$. Then, we pick an arbitrary location $i$ and insert the substring $\bfx_{i,L-1}$ at some location $j \neq i$. Then, since there are $2^{n-L+1}$ ways to select $\bfx$, and at most $\binom{n-L+1}{2} \leq \frac{(n-L+1)^2}{2}$ ways to pick the indices $i,j$, it follows that there are at most $2^{n-L+1} \cdot (n-L+1)^2/2$ ways to form a string which does not lie in $\cU(n,L)$. This in turn implies that
$$ |\cU(n,L)| \geq 2^n - (n-L+1)^2 \cdot 2^{n-L}.$$

For the upper bound, we observe that every word in $\cU(n,L)$ has to avoid concatenations of repeated substrings of length $L-1$.
As a result,
\begin{align*}
|\cU(n,L)| &\leq 2^{L-1} \cdot (2^{L-1} - 1) \cdot \prod_{j=1}^{\frac{n}{L-1}-2}  (2^{L-1} - j \cdot (L-1))\\
&\leq \left( 2^{L-1} \right)^{\frac{n}{2(L-1)}} \cdot \left( 2^{L-1} - \left(\frac{n}{2(L-1)} - 1\right) \cdot (L-1) \right)^{\frac{n}{2(L-1)}} \\
&\leq 2^n \cdot \left( 1 - (\frac{n}{2(L-1)} - 1) \cdot \frac{(L-1)}{2^{(L-1)}} \right)^{\frac{n}{2(L-1)}} \\
&\leq 2^n \cdot \exp \left( - \frac{n}{2^{L}} \cdot (\frac{n}{2(L-1)} - 1) \right),
\end{align*}
where the first inequality follows from upper bounding the product using an appropriate power of its first term only.

\section{The Repeat Replacement Encoder for $L > 2\log n + 4$ }\label{app:encLG}

The only difference between the repeat replacement encoder for $L > 2\log n + 4$ and the one for $L = 2 \log n + 4$ is a modification of Step 2, described below. There, we append the string $(1,1,\ldots,1,B(i), B(j),0,1)$ rather than the string $(B(i), B(j),0,1)$ to the current string estimate, where the length of the prepended string of $1$s equals $L - (2 \log n +4 )$.

\begin{algorithm}[H]
\caption{Repeat replacement encoder $\cE_{RR}$ for generating $(L-1)$-substring unique strings.}
\begin{algorithmic}[1]
\STATE If $\bfx_I$ is $(L-1)$-substring unique, set $\bfx = \bfx_I$ and STOP. Otherwise, set $\bfx^{(0)} = \bfx_I$, and 
let $k=1$.
  \STATE Suppose that $\bfx^{(k-1)}$ has a repeat at positions $(i,j)$ of length $L-1$. Let $\bfx^{(k)}$ be obtained by 
  deleting $\bfx^{(k-1)}_{j,L-1}$ from $\bfx^{(k-1)}$ and subsequently appending the string $(1,1,\ldots,1,B(i), B(j),0,1) \in \{0,1\}^{L-2}$ to the modified string $\bfx^{(k-1)}$.
  \STATE If $x^{(k)}_{L-1}=0$, i.e., if the $(L-1)\,$-st bit of $\bfx^{(k)}$ equals $0$, reset $x^{(k)}_{L-1}=1$ and update the 
  last two bits of $\bfx^{(k)}$ to $(1,1)$. If $\bfx^{(k)}$ is $(L-1)$-substring unique, set $\bfx= \bfx^{(k)}$ and STOP. Otherwise, set $k=k+1$, and go to Step 2.
\end{algorithmic}
\end{algorithm}

\section{The Primal Repeat Replacement Encoder for $L > \log n + 2 \log \log n + 8$}\label{app:encLSB}

Once again, the only difference between the encoder used for $L = \log n + 2 \log \log n + 8$ and the encoder used for $L > \log n + 2 \log \log n + 8$ is in one single step, Step 2. There, the string $(0,0,\ldots,0,1,B_r(i),1,1,\ldots,1)$ is inserted instead of the string $(0,0,\ldots,0,1,B_r(i),1),$ where the appended string of all $1$s has length $(L-2) - (\log n + 2\log \log n + 5)$.

\begin{algorithm}[H]
\caption{Primal repeat replacement encoder $\cE_{Prr}$ for generating $(L-1)$-substring unique strings.}
\begin{algorithmic}[1]
\STATE If $\bfx_I$ is $(L-1)$-substring unique, then set $\bfx = \bfx_I$ and STOP. Otherwise, set $\bfx^{(0)} = \bfx_I$, and let $k=1$.
  \STATE Suppose that $\bfx^{(k-1)}$ has a repeat at $(i,j)$ of length $L-1,$ where $(i,j)$ is primal. Generate $\bfx^{(k)}$ by replacing $\bfx^{(k-1)}_{j,L-1}$ with the string $(0,0,\ldots,0,1,B_r(i),1,1,\ldots,1)$ of length $L-5$.
  \STATE If $x_{L-1}^{(k)} = 0$, update $x_{L-1}^{(k)} = 1$ and append $(1,1,1)$ to $\bfx^{(k)}$. Otherwise, append $(1,0,1)$ to $\bfx^{(k)}$. If $\bfx^{(k)}$ is $(L-1)$-substring unique, set $\bfx = \bfx^{(k)}$ and STOP. Otherwise, set $k=k+1$, and go to Step 2.
\end{algorithmic}
\end{algorithm}

\end{appendices}
 
\end{document}